\documentclass[pdflatex,sn-mathphys-ay]{sn-jnl}% Math and Physical Sciences Author Year Reference Style
\usepackage{graphicx}%
\usepackage{multirow}%
\usepackage{amsmath,amssymb,amsfonts}%
\usepackage{amsthm}%
\usepackage{mathrsfs}%
\usepackage[title]{appendix}%
\usepackage{xcolor}%
\usepackage{textcomp}%
\usepackage{manyfoot}%
\usepackage{booktabs}%
\usepackage{syntax}
\usepackage{algpseudocode}%
\usepackage{listings}%
\usepackage{enumitem}
\usepackage[linesnumbered,ruled,vlined]{algorithm2e}
\usepackage{tcolorbox}
\setcitestyle{aysep={}}
\usepackage[utf8]{inputenc}
\usepackage{booktabs}
\usepackage{array} 

\makeatletter
\patchcmd{\NAT@citex}
  {\@citea\NAT@hyper@{%
     \NAT@nmfmt{\NAT@nm}%
     \hyper@natlinkbreak{\NAT@aysep\NAT@spacechar}{\@citeb\@extra@b@citeb}%
     \NAT@date}}
  {\@citea\NAT@nmfmt{\NAT@nm}%
   \NAT@aysep\NAT@spacechar\NAT@hyper@{\NAT@date}}{}{}

\patchcmd{\NAT@citex}
  {\@citea\NAT@hyper@{%
     \NAT@nmfmt{\NAT@nm}%
     \hyper@natlinkbreak{\NAT@spacechar\NAT@@open\if*#1*\else#1\NAT@spacechar\fi}%
       {\@citeb\@extra@b@citeb}%
     \NAT@date}}
  {\@citea\NAT@nmfmt{\NAT@nm}%
   \NAT@spacechar\NAT@@open\if*#1*\else#1\NAT@spacechar\fi\NAT@hyper@{\NAT@date}}
  {}{}

\makeatother

\theoremstyle{thmstyleone}%
\theoremstyle{thmstyletwo}%

\theoremstyle{thmstylethree}%

\begin{document}

\title[Article Title]{Two Birds One Stone: Effective Static Detection of Resource and Communication Deadlocks in Rust Programs}

\author[1]{\fnm{Yu} \sur{Zhang}}\email{2230766@tongji.edu.cn}
\author[1]{\fnm{Kaiwen} \sur{Zhang}}\email{2210133@tongji.edu.cn}
\author*[1]{\fnm{Guanjun} \sur{Liu}}\email{liuguanjun@tongji.edu.cn}
\author[2]{\fnm{Yuandao} \sur{Cai}}\email{ycaibb@cse.ust.hk}
\author[3]{\fnm{Shengchao} \sur{Qin}}\email{shengchao.qin@gmail.com}

\affil*[1]{\orgname{Tongji University}, \orgaddress{\city{Shanghai}, \country{China}}}

\affil[2]{\orgdiv{Hong Kong University of Science and Technology}, \orgaddress{\city{Hong Kong},\country{China}}}

\affil[3]{\orgdiv{Xidian University}, \orgaddress{\city{Xi’an},\country{China}}}

\abstract{
Although Rust claims to enable fearless concurrency, concurrency errors remain prevalent in practical applications, with deadlocks particularly prominent and harmful. 
Current static deadlock detection techniques either do not cater to Rust programs without incorporating Rust semantics or solely concentrate on resource deadlocks, disregarding communication deadlocks.
In this paper, we present \textsc{RcChecker}, the first static detection method that simultaneously targets resource and communication deadlocks for Rust. 
At its core, \textsc{RcChecker} introduces the signal-lock graph to characterize the interactions between threads through lock variables and condition variables. 
The construction of the signal-lock graph involves utilizing flow-sensitive and context-sensitive analyses to track variable lifetime details and abstracting actions associated with condition variables into signal resource holdings and requests.
By examining cycles within the signal-lock graph, we can effectively identify resource and communication deadlocks in a unified way. 
Compared to the existing Rust deadlock detection tool, \textsc{RcChecker} effectively eliminates eight resource deadlock false positives and detects seventeen errors that the existing tool fails to identify.
Furthermore, \textsc{RcChecker} reports seven previously undiscovered issues in real-world Rust applications, including two resource deadlocks and five communication deadlocks.
}




\maketitle

\section{Introduction}
\label{Intro}

Rust \citep{rustPL} is an emerging programming language that excels in constructing efficient and secure low-level software. Inspired by C, Rust inherits excellent runtime performance and addresses safety concerns through rigorous compile-time checks. 
Due to its unique combination of performance and safety features, Rust is gaining increasing popularity among developers. Rust implements a unique safety mechanism based on ownership rules to ensure both memory safety and thread safety. However, Rust programs are not entirely immune to security vulnerabilities, particularly those related to concurrency. Studies \citep{qin2024understanding, Hu2022comprehensiveness, yu2019fearless} indicate that over 50\% of concurrency vulnerabilities in Rust are related to deadlocks. This paper focuses on Rust programs, primarily addressing \textit{resource deadlocks}, where a thread attempts to reacquire a non-reentrant lock or cyclic waiting for locks occurs between threads, and \textit{communication deadlocks related to condition variables}, which arise from lost notify signals and cyclic waiting involving condition variables and locks.

Program analysis is the predominant method for deadlock detection. Dynamic analysis \citep{ cai2020low,cai2014magiclock,cai2012magicfuzzer,eslamimehr2014sherlock,samak2014trace,sorrentino2015picklock,zhou2017undead}  identifies potential deadlocks by monitoring program behavior at runtime, but it suffers from limited coverage. In contrast, static analysis \citep{cai2022peahen,naik2009effective,brotherston2021compositional,engler2003racerx,kroening2016sound} inspects program source code to detect potential deadlocks, effectively exploring obscure program paths that are difficult to reach during execution.
Despite extensive research in deadlock detection, we note that most studies predominantly focus on resource deadlocks in , with very few works exploring both resource and communication deadlocks \citep{hovemeyer2004finding,joshi2010effective,zhou2022deadlock}.

Specifically, existing works for Java, C, and C++ cannot be easily extended to Rust. First, locks in Rust are strictly managed via RAII (Resource Acquisition Is Initialization) through guard types, which lack explicit unlock methods. The lock release is implicitly triggered when the guard lifetime ends (e.g., at scope exit), depending on compiler-injected drop logic. Traditional approaches that rely on explicit unlock function pairs or destructor analysis are hard to model lock states bound to lifetimes in Rust. 
Additionally, the ownership rules in Rust enforce concurrent data sharing through reference-counted smart pointers, requiring deadlock detection tools to trace ownership transfers and smart pointer management.

Additionally, there are only a few existing works on deadlock detection for Rust, and current approaches face specific challenges.
\textsc{Stuck-me-not} \citep{ning2020stuck} focuses solely on identifying double locks (i.e., a thread attempts to reacquire a non-reentrant lock) in the blockchain software. \textsc{Lockbud} \citep{qin2024understanding, Boqin2020understanding,lockbud} for resource deadlocks uses only type information in parameters (ignoring the alias information) to guide heuristic analysis for inter-procedural methods, which suffers from numerous false positives and false negatives. 

To address the issues mentioned above, in this work, we present \textsc{RcChecker}, the first static method for large-scale Rust programs that detects both resource and communication deadlocks in a unified way.
The key of \textsc{RcChecker} is the inclusion of the new \textit{signal-lock graph}, a novel structure that characterizes the interactions between threads through both lock variables and condition variables, where each node represents a scoped lock or a condition variable, and the edges indicate the dependency and alias relationships between the variables.
The construction of the signal-lock graph is achieved by employing flow-sensitive and context-sensitive analysis, effectively tracking variable lifetime information and abstracting operations related to condition variables into signal resource holdings and requests. 

\textsc{RcChecker} implements the detection of four types of deadlocks: \textit{double lock} caused by reacquiring a non-reentrant lock within, \textit{conflict lock} resulting from improper interactions between locks across threads, \textit{conflict signal-lock} due to improper interactions between locks and condition variables across threads, and \textit{lost notification} arising from the incorrect use of condition variables. 
\textsc{RcChecker} conducts static deadlock detection in three stages. First, it performs an Andersen-style pointer analysis specifically for Rust programs, introducing two new rules based on the ownership model and smart pointer mechanisms to optimize the process. 
Next, \textsc{RcChecker} performs a flow-sensitive and context-sensitive analysis using points-to and lifetime information to construct a signal-lock graph, where each cycle indicates a potential deadlock. 
Finally, drawing on four distinct types of deadlocks, four cycle patterns are defined in the signal-lock graph to help to indicate resource deadlocks and communication deadlocks.
 
To evaluate the detection effectiveness of \textsc{RcChecker}, we first conduct a comparative experiment with the existing tool. \textsc{RcChecker} identifies 79 deadlocks across 18 real-world programs, with the detection time accounting for approximately 15.47\% of the project build time. Although it introduces an additional overhead of about 8.31\% compared to existing tools, \textsc{RcChecker} eliminates 8 false positives in the existing tool, improving detection precision. We further analyze the causes of false positives in existing tools and categorize them into three main types.
Subsequently, to verify the capability of \textsc{RcChecker} in detecting both communication deadlocks and resource deadlocks, we test it on 11 programs and identify 17 bugs that are undetectable by the existing tool. Among them, 7 newly reported deadlocks originate from real-world systems, including 2 resource deadlocks and 5 communication deadlocks.

To sum up, this paper makes the following contributions:
\begin{itemize}
    \item We propose \textsc{RcChecker}, the first static method for Rust programs that detects both resource and communication deadlocks in a unified way. 
    \item \textsc{RcChecker} reports seven previously unreported deadlocks in real-world systems, including two resource deadlocks and five communication deadlocks.
\end{itemize}

This paper is organized as follows: Section 2 introduces the features of the Rust programming language and the types of deadlocks detected by \textsc{RcChecker}. Section 3 provides basic symbols and relevant definitions. Section 4 details the implementation of RcChecker. Section 5 evaluates the performance of RcChecker. Sections 6 and 7 discuss the limitations of the current work and review related research. Finally, Section 8 concludes this paper and outlines future research directions.

\section{Background} \label{section2}
\begin{figure*}[t] 
  \centering
   \includegraphics[width=\linewidth]{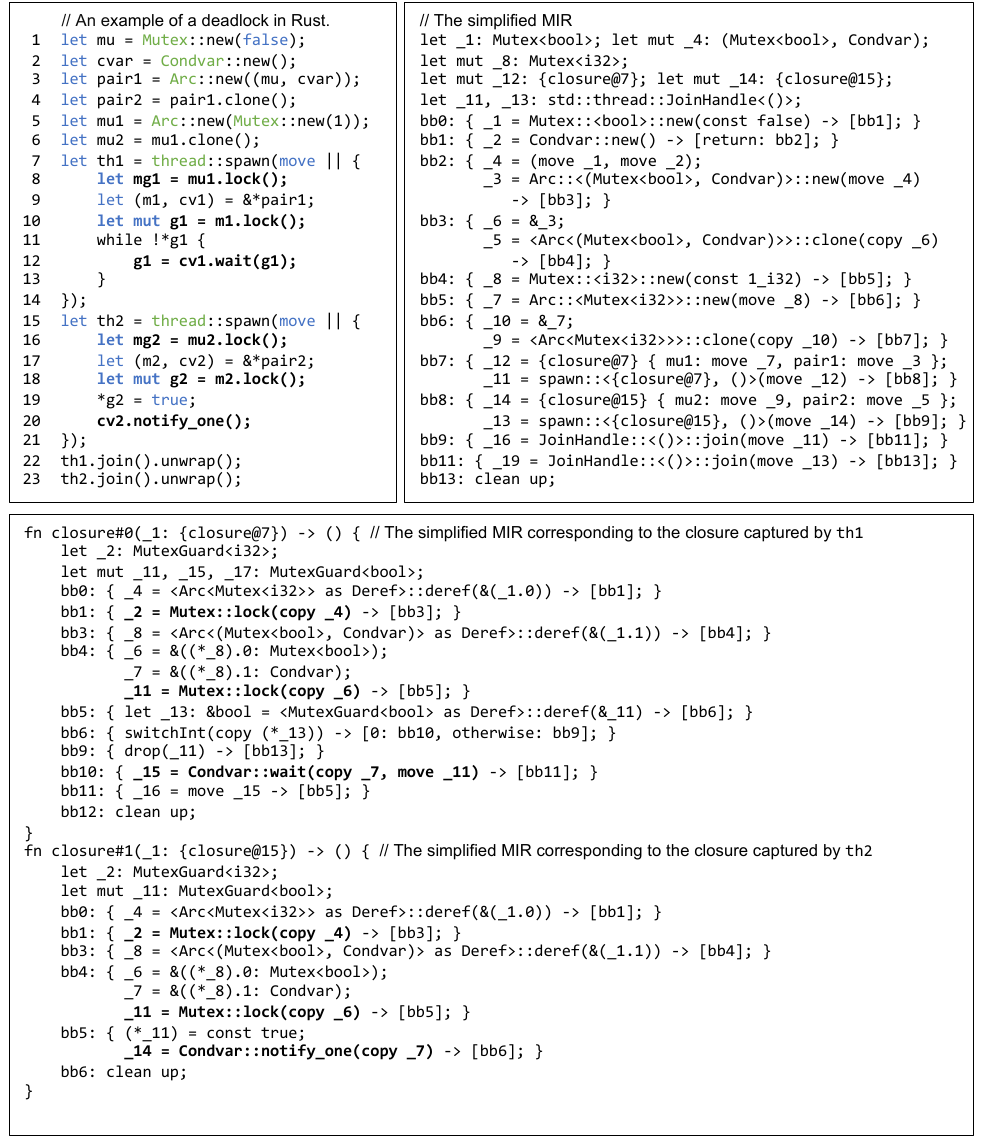}
    \caption{A Rust code snippet that can cause a deadlock and its simplified MIR}
    \label{bg}
\end{figure*}

In this section, we introduce some background knowledge through the code snippet shown in Figure \ref{bg}, including the ownership model and smart pointers in Rust, Mid-level Intermediate Representation, thread synchronization mechanisms, and the types of deadlocks detectable by \textsc{RcChecker}.

\subsection{Ownership and Smart Pointer}
The core of safety mechanisms in Rust is \textit{Ownership}, where each value can have only one \textit{owner}, and when the owner goes out of scope, the value is dropped. 
When assigning a value (e.g., \verb|let y = x|) or passing function arguments by value (e.g., \texttt{foo(x: T)}), the ownership of the resource is transferred, a process known as a \textit{move}.
The Rust compiler guarantees that a variable cannot be accessed after its ownership is moved, preventing dangling references and ensuring memory safety. 
For example, in the Rust code shown in Figure \ref{bg}, after executing \texttt{pair1 = Arc::new((mu, cvar))} in line 3, the ownership of the values held by \texttt{mu} and \texttt{cvar} is transferred to the newly created \texttt{Arc} instance \texttt{pair1}. As a result,\texttt{mu} and \texttt{cvar} can no longer be accessed directly.

Rust also supports temporary access to data without transferring ownership, known as \textit{borrowing}. 
This is achieved by passing a reference to the value, either as an immutable reference (e.g., \texttt{\&T}) or a mutable reference (e.g., \texttt{\&mut T}). An immutable reference allows read-only access to the underlying data (i.e., the value of type \texttt{T}), while a mutable reference allows both read and write access to the underlying data.
For example, a function can borrow data by accepting a reference as an argument for immutable borrowing (e.g., (\texttt{foo(x: \&T)}) or for mutable borrowing (e.g., \texttt{foo(x: \&mut T)}). Rust enforces strict borrowing rules: at any given time, either one mutable reference or multiple immutable references can exist.
The \textit{lifetime} of a variable begins when it is created and ends when it is destroyed. When a variable has existing references, its ownership cannot be transferred. The Rust compiler uses a borrow checker to verify the lifetimes of references at compile time. The borrow checker ensures that the lifetime of a reference does not exceed the lifetime of the data it points to, thereby preventing dangling references. Additionally, it enforces borrowing rules by tracking the scope and validity of references, ensuring the safety and correctness of references throughout their lifetimes.

Smart pointers are data structures that act like pointers but also carry additional metadata and capabilities. In Rust, smart pointers provide more complex memory management and ownership control compared to regular references.
This paper primarily focuses on \textit{reference counting smart} \textit{pointers} in Rust, which manages memory by implementing shared ownership through tracking the number of references to data. \verb|Rc<T>| facilitates multiple ownership in single-threaded environments, whereas \verb|Arc<T>| is designed for multithreaded environments, using atomic operations to update the reference count and ensure safety during concurrent access. 
For example, in line 3 of the Rust code shown in Figure \ref{bg}, \texttt{Arc::new((mu, cvar))} creates a new \texttt{Arc} instance \texttt{pair1}. In line 4, \texttt{pair1} creates \texttt{pair2} through the \texttt{clone()} method, which is a new \texttt{Arc} instance that shares ownership of \texttt{(mu, cvar)} with \texttt{pair1}. The \texttt{clone()} function atomically increments the reference count, ensuring safe data sharing in a multi-threaded environment. Threads \texttt{th1} and \texttt{th2} can thus safely share ownership of \texttt{(mu, cvar)}.

\subsection{Mid-level Intermediate Representation}
Static analysis \citep{li2021mirchecker,bae2021rudra,vallee2010soot,sui2016svf,tan2022tai-e,li2024context} is predominantly based on intermediate representations (IRs), which provide a compact and uniform format that facilitates analysis and optimization. 
The Rust compiler can generate several intermediate representations, including HIR (High-level Intermediate Representation), MIR (Mid-level IR), and LLVM IR (Low-Level Virtual Machine Intermediate Representation) \citep{lattner2004llvm}.

\begin{table}[t]
\caption{Core syntax of the MIR model}
\label{MIR model}
\setlength{\tabcolsep}{5pt}

\begin{tabular}{llcl}
\toprule
\textbf{Constant} & 
  $c$& \\
\textbf{Terminator} & 
  $te$ & $::=$ & $\texttt{Goto}(bb) \mid \texttt{SwitchInt}(op, bb_1, bb_2) \mid  \texttt{Drop}$ \\
\textbf{Local} & 
  $o $&$\in$&$ \{o_1, o_2, ...\}$ \\
\textbf{Place} & 
  $p$ & $::=$ & $o \mid *p \mid p.x$ \\
  
\textbf{Operand} & 
  $op$ & $::=$ & $c \mid \texttt{(copy)}\,p \mid \texttt{move}\,p$ \\

\textbf{Rvalue} & 
  $rv$ & $::=$ & $\&p \mid op$ \\

\textbf{Statement} & 
  $s$ & $::=$ & $s_1; s_2 \mid p = rv \mid p = \texttt{func}(op_1, op_2, ...)$ \\
\textbf{Basic Block} & 
  $b$ & $::=$ & $\texttt{bb}_i \{s_1; s_2; ...; te\}$ \\
\textbf{Instance} & 
  $\gamma$ & $::=$ & $\texttt{func}([rv_1, rv_2, ...], p_{ret}) \{b_1, b_2, ...\}$ \\
\bottomrule
\end{tabular}
\end{table}

\textsc{RcChecker} relies on MIR for static analysis because it includes both control flow and lifetime information of variables, which are essential for detecting deadlocks. Table \ref{MIR model} introduces the core MIR syntax used in this paper. We illustrate these concepts by analyzing the simplified MIR\footnote{To improve readability, the MIR shown here has been simplified by removing non-essential details like cleanup operations, error-handling branches, and unnecessary basic blocks. The full MIR includes additional low-level details such as memory management and lifetime annotations.} corresponding to the Rust code in Figure \ref{bg}.

The Rust compiler achieves type safety and performance optimization through \textit{monomorphization}, a mechanism that generates specialized implementation instances for generic constructs (functions, types, closures) based on specific type parameters. For example, the generic function \texttt{Arc::new(T)} produces distinct implementation instances for each concrete type \texttt{T} used in the program, such as \texttt{(Mutex<bool>, Condvar)} in line 3 and \texttt{Mutex<i32>} in line 5. 
Closures follow the same rule, with each unique closure type manifesting as an independent instance at the MIR level. Specifically, the closure \texttt{th1} (capturing \texttt{mu1} and \texttt{pair1}) creates a concrete \texttt{closure\#0} instance, while the closure \texttt{th2} (capturing \texttt{mu2} and \texttt{pair2}) produces a distinct \texttt{closure\#1} instance. 
Non-generic functions map directly to a single implementation instance without undergoing monomorphization, as they require no type-specific specialization.

The core structure of MIR is organized around basic blocks, which serve as the fundamental units of control flow. 
Each basic block $b$ consists of a sequence of statements and a terminator. 
A statement $s$ performs specific operations, while a terminator $te$ dictates control-flow transitions. For instance, in the instance \texttt{closure\#0}, the terminator \texttt{switchInt} in \texttt{bb6} evaluates a boolean condition to determine the next control flow path, thereby forming a loop structure.
A local $o$ represents an abstract storage unit (e.g., \texttt{_1: Mutex}), serving as a variable in the MIR representation. 
A place $p$ describes the memory location where values are stored, including direct access to locals, projections such as struct fields obtained via dereferencing (e.g., \texttt{(*_8).0} in \texttt{closure\#0}, which first dereferences an \texttt{Arc} and then accesses the first element of its inner tuple), or indexed elements of an array.
An operand $op$ represents a concrete value used in computations, which may be a constant (e.g., \texttt{const false}) or a value obtained through \texttt{move} and \texttt{copy} semantics. At the MIR level, explicit \texttt{copy} markers may be omitted depending on the type implementation of the \texttt{Copy} trait and compiler optimizations.
An rvalue $rv$ represents the result of operations performed on operands, which includes references to places or the direct use of operands.
An instance $\gamma$ denotes the concrete implementation of a function or closure, where $[rv_1, rv_2, \dots]$ represents its parameter list, and $p_{ret}$ denotes its return value.

\subsection{Synchronization Mechanism of Rust} \label{tsS}
This section explores thread synchronization mechanisms in Rust, focusing on thread creation, the use of locks, and the application of condition variables.

\textbf{Thread Creation and Completion}. In Rust, a thread is created using \verb|thread::spawn()| with a closure that contains the code to be executed in the new thread. 
The \verb|move| keyword is typically used to transfer ownership of the closure to the new thread. 
If a value is needed in the main thread, it should be shared using references or \verb|Arc<T>|.
For instance, in Figure \ref{bg}, \texttt{th1} and \texttt{th2} are created in lines 7 and 15, with the \texttt{move} keyword transferring ownership of the captured variables \texttt{mu1}, \texttt{pair1}, \texttt{mu2}, and \texttt{pair2} to the respective threads.
\verb|thread::join()| waits for a thread to finish and retrieves its return value. It is a method on the \verb|JoinHandle| type, which is the handle returned by \verb|thread::spawn()| to manage the thread lifetime.

\textbf{Lock.} The most commonly used locks are \verb|Mutex| and \verb|RwLock|. \verb|Mutex| is a non-reentrant lock, meaning the same thread cannot lock it multiple times without unlocking. \verb|RwLock| is used when multiple readers are allowed, but only one writer can hold the lock at a time.
Unlike traditional languages such as Java and C, locks in Rust protect data rather than code segments and do not provide explicit unlock functions.
Rust employs the RAII (Resource Acquisition Is Initialization) paradigm to implement scoped lock management. 
For example, with \verb|Mutex|, its guard object is \verb|MutexGuard|, which provides access to the protected data while holding the lock. 
The Rust Compiler relies on the lifetime of the \verb|MutexGuard| to manage the lock, which is automatically released when the \verb|MutexGuard| goes out of scope.
We must obtain the \verb|MutexGuard| by invoking the \verb|Mutex::lock()| function to access the data\footnote{Rust uses a strategy called \textit{poisoning} to handle thread panics while holding a lock. When a thread panics, the lock is marked as \textit{poisoned}, and subsequent attempts to acquire it return an \texttt{Err} result, helping detect potential data corruption. In most cases, \texttt{unwrap()} is used on the \texttt{Result} to obtain the \texttt{MutexGuard}, assuming poisoning is not an issue. The poisoning process is omitted in this paper since it is not directly relevant to deadlock detection.}. Similarly, \verb|RwLock| obtains the \verb|RwLockReadGuard| through \verb|RwLock::read()| and the \verb|RwLockWriteGuard| through \verb|RwLock::write()|.

In Figure \ref{bg}, the \texttt{mutex} \texttt{mu1} protects the data \texttt{i32} with the value \texttt{1}. When accessing this data, \texttt{mu1} acquires a \texttt{MutexGuard} (i.e., \texttt{mg1}) through the \texttt{lock()} method (line 8). From the moment the lock is acquired until the end of thread \texttt{th1} (line 14), \texttt{mu1} remains in a locked state. During this period, other threads cannot access the data protected by \texttt{mu1} through the same \texttt{MutexGuard}, ensuring thread safety. In Rust, locks are typically implemented using smart pointers. For example, \texttt{mu1} is wrapped in an \texttt{Arc}, allowing other threads to access the data by acquiring their own \texttt{MutexGuard} through a cloned \texttt{Arc} instance (e.g., \texttt{th2} accesses the data through \texttt{mu2}). However, the lock mechanism ensures that only one thread can hold the \texttt{MutexGuard} at a time, preventing concurrent access to the protected data.

\textbf{Condition Variable.}  
Condition variables represent the ability to block a thread such that it consumes no CPU time while waiting for an event to occur.
Condition variables are typically associated with a boolean predicate (a condition) and a lock. 
The predicate is always verified inside of the \verb|Mutex| before determining that a thread must be blocked. 
\verb|Condvar::wait(MutexGuard)| releases the associated \verb|Mutex| and blocks the current thread until the condition variable receives a notification. 
When the thread is notified, it automatically reacquires the lock. \verb|Condvar::notify_one()/notify_all()| methods wake up threads waiting on the condition variable. 
For example, in Figure \ref{bg}, a condition variable \texttt{cvar} is used to synchronize threads \texttt{th1} and \texttt{th2}. The boolean predicate \texttt{*g1} is protected by a \texttt{Mutex} and verified inside a \texttt{while} loop in \texttt{th1}. If the predicate is \texttt{false}, \texttt{th1} releases the lock and blocks itself by calling \texttt{cv1.wait(g1)}. Meanwhile, \texttt{th2} acquires the lock, updates the predicate to \texttt{true}, and notifies \texttt{th1} by calling \texttt{cv2.notify\_one()}. Upon receiving the notification, \texttt{th1} reacquires the \texttt{Mutex} and exits the loop. 

When using condition variables, it is prudent to enclose the wait operation within a \texttt{while} loop to ensure robust synchronization. 
If the wait operation is placed as a standalone statement or within a branching structure rather than a loop structure, it can result in spurious wake-ups and lost notifications due to the condition not being evaluated, compromising the reliability of concurrent processes and increasing the risk of deadlocks. 
Additionally, failing to correctly update the predicate before the notify operation can result in permanent blocking.
\begin{figure}[t] 
  \centering
  \includegraphics[width= \linewidth]{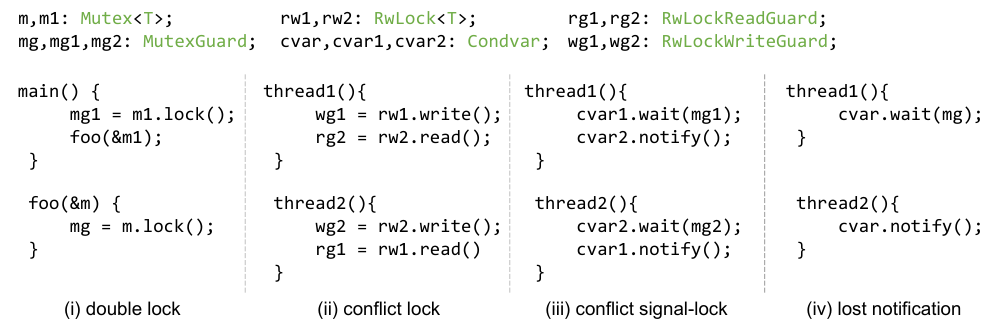}
  \caption{Types of resource deadlocks and communication deadlocks detected by \textsc{RcChecker}}
  \label{pattern}
\end{figure}

\subsection{Types of Deadlocks detected by \textsc{RcChecker}}
\label{typeS}
\textsc{RcChecker} can detect resource deadlocks and communication deadlocks, which can be further classified into four types: double lock, conflict lock, conflict signal lock, and lost notification. Figure \ref{pattern} provides a summary of these patterns through simplified Rust code snippets. In the following sections, we elaborate on the four types in detail. 

\textbf{Double lock.} A deadlock occurs when a thread attempts to reacquire a non-reentrant lock, a type of resource deadlock. 
Figure \ref{pattern}(I) illustrates a double lock involving function calls and references passed as arguments. When calling \verb|foo()|, the lock on \verb|mg1| has not yet been released, and attempting to lock the same resource again within \verb|foo()| results in a deadlock.

Although double locking may appear straightforward, the associated issues can become complex and challenging to identify when function calls and pattern matching are involved.
Insufficient understanding of the lifetime and contextual scope of locks can also lead to confusion. 
For example, in the \texttt{main()} function of Figure \ref{pattern}(I), if \texttt{mg1 = m1.lock()} is changed to \texttt{m1.lock()}, acquiring the lock in \texttt{foo()} does not result in a deadlock. 
This is because \texttt{mg1 = m1.lock()} binds the \texttt{MutexGuard} to the variable \texttt{mg1}, causing the lock to remain held until \texttt{mg1} goes out of scope, whereas \texttt{m1.lock()} does not bind the \texttt{MutexGuard}, causing it to be released immediately after the statement ends. As a result, reacquiring the lock does not cause a deadlock.

\textbf{Conflict lock.} A deadlock occurs when threads cyclically wait for locks due to conflicting lock acquisition orders, a type of resource deadlock. 
Figure \ref{pattern}(II) illustrates a classic example of the conflict lock. 
Threads, \texttt{thread1()} and \texttt{thread2()}, acquire locks on \texttt{rw1} and \texttt{rw2} in different orders. If \texttt{thread1} holds \texttt{rw1} and requests \texttt{rw2}, while \texttt{thread2} holds \texttt{rw2} and requests \texttt{rw1}, a mutual waiting situation occurs, leading to a deadlock.

\textbf{Conflict signal-lock.} Improper interactions between condition variables and locks or among the condition variables can result in communication deadlock. It is similar to the conflict lock, except that the cyclic waiting involves condition variables. 
Figure \ref{pattern}(III) illustrates a common scenario of mutual waiting for notification signals. In this case, \texttt{thread1} waits for \texttt{thread2} to send a notification via \texttt{cvar2.notify()}, while \texttt{thread2} waits for \texttt{thread1} to send a notification via \texttt{cvar1.notify()}. This mutual waiting can lead to a deadlock if neither thread proceeds to send the required notification.

Similarly, this mutual waiting situation can also involve both locks and condition variables. Figure \ref{bg} is a typical example: while thread \texttt{th1} holds \texttt{mu1} and enters a waiting state for th2 to send a notification signal, \texttt{th2} needs to acquire \texttt{mu2} first before it can send the notification. Since \texttt{mu1} and \texttt{mu2} are the same lock shared through \texttt{Arc}, they cannot be accessed simultaneously, resulting in a mutual waiting situation.

\begin{figure}[t] 
  \centering
  \includegraphics[width= \linewidth]{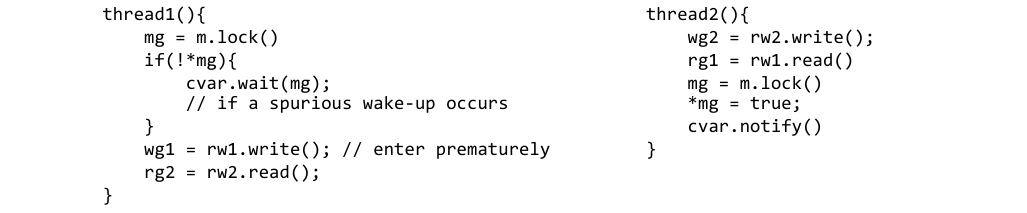}
  \caption{An example of a spurious wakeup indirectly leading to deadlock}
  \label{spurious}
\end{figure}
\textbf{Lost notification.} When a thread executing the wait operation fails to receive the notification signal for some reason, resulting in indefinite waiting, this forms a communication deadlock. 
The simplest cause of indefinite waiting is the absence of the necessary \texttt{notify} statement in the program itself. 
Furthermore, as discussed in Section 2.3 regarding the usage specifications of condition variables, placing the \texttt{wait} operation within a loop structure for conditional checking represents a more robust implementation approach. Whether it is a standalone wait statement or its usage within a branch condition (such as an \texttt{if} statement), we categorize both cases as instances of lost notifications.

For a standalone wait statement, this scenario is straightforward to understand, as illustrated in Figure \ref{pattern}(IV). During concurrent thread execution, it is possible for \verb|thread2| to send a notification before the wait operation, leading to \verb|thread1| waiting indefinitely.
Placing the wait statement within a branch condition may lead to spurious wakeups. Although spurious wakeups do not directly cause deadlocks, they can cause threads to resume execution prematurely before the condition is met, potentially allowing concurrent code segments that should be mutually exclusive to run simultaneously, thereby indirectly precipitating deadlocks and other concurrency issues. 
As illustrated in Figure \ref{spurious}, this example demonstrates a conflicting access order between \texttt{thread1} and \texttt{thread2} when accessing shared resources \texttt{rw1} and \texttt{rw2}, with an attempt to coordinate their access using a condition variable \texttt{cvar}. When the condition variable is used correctly (by verifying the condition predicate within a loop structure before executing \texttt{wait}), it ensures that \texttt{thread1} accesses \texttt{rw1} and \texttt{rw2} only after \texttt{thread2} has completed, thereby preventing deadlock. However, if \texttt{wait} is placed within an \texttt{if} condition, \texttt{thread1} may be spuriously awakened without rechecking the predicate, leading to premature access of \texttt{rw1} and \texttt{rw2}. If \texttt{thread2} executes concurrently during this time, deadlock may occur.
Since spurious wakeups are equivalent to failing to receive the proper and valid notification signal, we categorize them as lost notification situations and include them within the scope of unified analysis.

\section{Preliminary} \label{preS}
In this section, we introduce the foundational symbols and relevant definitions used in our analysis method.

\begin{table}[t]
\caption{Basic symbols}
\centering
\renewcommand{\arraystretch}{1.1}  
\begin{tabular}{llll}
\toprule
\textbf{Basic Block} &  $b $ & $\in $  & $\mathcal{B}$ \\
\textbf{Statement} & $s $ & $\in $ & $\mathcal{S}$ \\
\textbf{Instance} & $\gamma $ & $ \in $ & $\Gamma$ \\
\textbf{Place} & $p $ & $\in $ & $\mathcal{P}$ \\
\textbf{Calling Context} & $\mathit{ctx}$ &&\\
\textbf{Location} & $l$&&\\
\textbf{Thread} & $th$ && \\
\textbf{Variable} & $v$ & $\in$ & $\mathcal{V} \subseteq \mathcal{P} \times \Gamma$ \\
\textbf{Points-to Set} & 
  $pts $ & $:$ & $\mathcal{V} \to 2^\mathcal{V}$ \\
\textbf{Operation Type} & $\tau $ & $\in $ & $\{\textsf{MX}, \textsf{RL}, \textsf{WL}, \textsf{WT}, \textsf{NT}\}$ \\
\textbf{Interaction Relationship Type} & $k $ & $\in $ & $\{\textsf{D}, \textsf{A}\}$ \\
\textbf{Variable node} & $n$ & $\in$ & $\mathcal{N}$ \\
\textbf{Alive Nodes Mapping} & 
  $\mathbb{A} $ & $:$ & $\mathcal{N} \to 2^{\mathcal{N}}$ \\
\textbf{CondVar Relation} &
  $\mathbb{R} $ & $\subseteq$ & $ \mathcal{N}  \times \mathcal{N}$ \\
\bottomrule
\end{tabular}
\label{symbol}
\end{table}

\textbf{Basic Symbols.}
Table \ref{symbol} summarizes the basic symbols and abstract domains used in this paper.
The symbols $\mathcal{B}$, $\mathcal{S}$, $\Gamma$, and $\mathcal{P}$ respectively denote the sets of basic blocks, statements, instances, and places. 
The symbols $ctx$, $l$, and $th$ record the calling context, the position (line number) in the source code, and the thread, respectively.
The thread $th$ is represented by a triple $(p_{jh}, ctx_{sp}, ctx_{jn})$, where $p_{jh}$ denotes the returned \texttt{joinHandle} variable, $ctx_{sp}$ represents the calling context of the \texttt{spawn()} function at the thread creation point and $ctx_{jn}$ signifies the context information of the \texttt{join()} function, with $ctx_{jn}$ marked as \texttt{null} if the thread does not explicitly call \texttt{join()}.
We let the symbol $v$ denote a scoped lock or condition variable, represented as a tuple $(p,\gamma)$.
The symbol $pts$ represents the points-to relationship.

We use the symbol $\tau$ to indicate the type of operations related to locks and condition variables. Specifically, the symbol \textsf{MX} denotes the operation \texttt{Mutex::lock()}; the symbols \textsf{RL} and \textsf{WL} denote the operations \texttt{RwLock::read()} and \texttt{RwLock::write()} respectively; and the symbols \textsf{WT} and \textsf{NT} represent the methods \texttt{Condvar::wait(MutexGuard)} and \texttt{Condvar::notify()}, respectively.
We use the symbol $k$ to denote the interaction relationship between the lock variables and the condition variables. 
When one variable is acquired while another is held, we consider a dependency relationship to exist between these two variables, denoted as $k = \textsf{D}$. For example, in lines 8 and 10 of Figure \ref{bg}, we define a dependency relationship between variables \texttt{mg1} and \texttt{g1}, indicating that \texttt{g1} is acquired while \texttt{mg1} is held. When an aliasing relationship exists between two variables, we denote it as $k = \textsf{A}$. For instance, in lines 8 and 9 of Figure \ref{bg}, variables \texttt{mg1} and \texttt{mg2} represent two acquisition operations on the same lock, thus establishing an aliasing relationship between them.

\begin{figure}[t] 
  \centering
    \includegraphics[width=\linewidth]{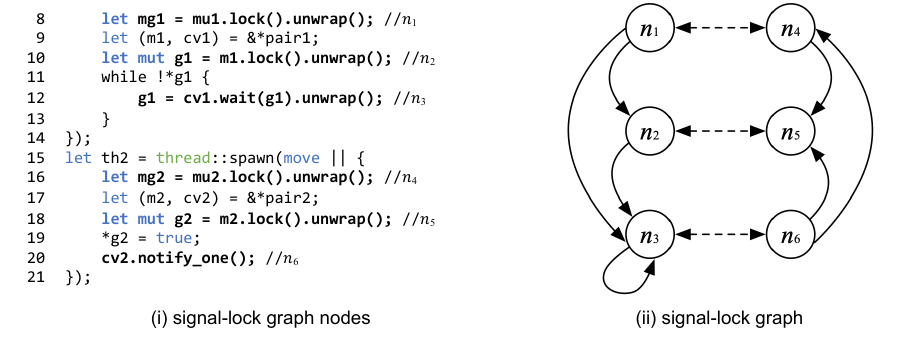}
    \caption{The node information, signal-lock graph, and dependency cycles of the deadlock example shown in Fig. \ref{bg}}
    \label{lg}
\end{figure}

We create a corresponding node structure $n$ for each lock and condition variable, represented as a tuple $(v, \tau, l, ctx, th)$.
Taking the code in Figure \ref{bg} as an example, we create corresponding nodes for the different operations on lock variables and condition variables. As shown in Figure \ref{lg}(i), a total of six nodes are created, labeled $n_1$ through $n_6$, corresponding to the statements at lines 8, 10, 12, 16, 18, and 20, respectively. 
The detailed construction process of the node structure is specifically described in Section \ref{slgS}.

Additionally, the symbol $\mathbb{A}$ is used to record the mapping of alive nodes. When a node is created, other nodes that have not yet reached the end of the lifetime (referred to as alive nodes) are recorded. For example, when node $n_3$ is created at line 12 of the code in Figure \ref{bg}, the variables corresponding to nodes $n_1$ and $n_2$ are still within the lifetime, hence $\mathbb{A}(n_3) = \{n_1, n_2\}$. The $\mathbb{A}$ mappings for the remaining nodes are shown in Figure \ref{lg}(i). The symbol $\mathbb{R}$ represents the relationship between a condition variable and its associated lock. Through the \texttt{Condvar::wait(MutexGuard)} statement, $\mathbb{R}$ records the node corresponding to the condition variable with the node corresponding to the scoped lock. For example, $\mathbb{R}(n_3) = n_2$. The recording and maintenance of $\mathbb{A}$ and $\mathbb{R}$ are explained in Section \ref{slgS}.

\textbf{Definition 1.} A \textit{signal-lock graph} (SLG) is a directed graph, denoted as $\mathcal{G}=(\mathcal{N}, \mathcal{E})$.
\begin{itemize}
    \item $\mathcal{N}$ is the set of all nodes $n$ corresponding to lock and condition variables. 
    \item $\mathcal{E}$ is a set of edges. Each $e = (n_i, n_j, k) \in \mathcal{E}$ represents an edge between $n_i$ and $n_j$. 
     $k = \textsf{D}$  indicates that the edge is a dependency edge, represented as a directed edge  $n_i \to n_j$.
      $k = \textsf{A}$  indicates that the edge is an alias edge, represented as a bidirectional edge  $n_i \leftrightarrow n_j$.
\end{itemize}
The signal-lock graph is utilized to depict the interaction relationships between lock variables and condition variables. The SLG corresponding to the code segment shown in Figure \ref{bg} is illustrated in Figure \ref{lg}(ii). The detailed construction process of the SLG is thoroughly discussed in Section \ref{slgS}.

\textbf{Definition 2.}
In the signal-lock graph, we define a cycle with at least one dependency edge as the \textit{dependency cycle}, denoted as $\mathcal{C} = (\mathcal{N}_c, \mathcal{E}_c)$, where $\mathcal{N}_c$ represents all the nodes and $\mathcal{E}_c$ represents all the edges in the dependency cycle.
To capture both intra-thread and inter-thread deadlocks, we classify the set of all dependency cycles (denoted as $\mathbb{C}$) into \textit{intra-thread dependency cycles} and \textit{inter-thread dependency cycles} based on the threads where the dependency cycle nodes are located.
In Figure \ref{lg}(ii), $n_1 \rightarrow n_3 \rightarrow n_6 \rightarrow n_5 \rightarrow n_1$ forms a dependency cycle. The deadlock detection approach based on dependency cycles is detailed in Section \ref{ddS}.

\textbf{Problem Statement.} After introducing the basic notions and relevant definitions, we summarize three problems we aim to address: 
\begin{enumerate}
    \item How to effectively analyze and handle variable aliasing relationships associated with the ownership mechanism and smart pointer rules in Rust.
    \item How to construct a signal-lock graph to characterize the interactions between threads through lock variables and condition variables.
    \item How to implement the detection of both resource deadlocks and communication deadlocks using the signal-lock graph.
\end{enumerate}

Section \ref{methodS} provides detailed solutions to the problems outlined above, summarized in the following three points.

\begin{enumerate}
    \item We implement the Andersen-style pointer analysis \citep{andersen1994program} specifically for Rust programs and propose two rules grounded in the ownership and smart pointer mechanisms to optimize the process. (Section \ref{paS})
    \item We analyze the interactions between threads through lock variables and condition variables by tracking the lifetimes in a flow-sensitive and context-sensitive analysis while abstracting the operations associated with condition variables into signal resource holdings and requests to construct the signal-lock graph. (Section \ref{slgS})
    \item We traverse the signal-lock graph to identify dependency cycles, which represent potential deadlocks. Based on the characteristics of the four summarized deadlock types, we analyze these dependency cycles to identify and report deadlocks. (Section \ref{ddS}) 
\end{enumerate}

\section{Methodology}  \label{methodS}
\begin{figure*}[t] 
  \centering
  \includegraphics[width= \linewidth]{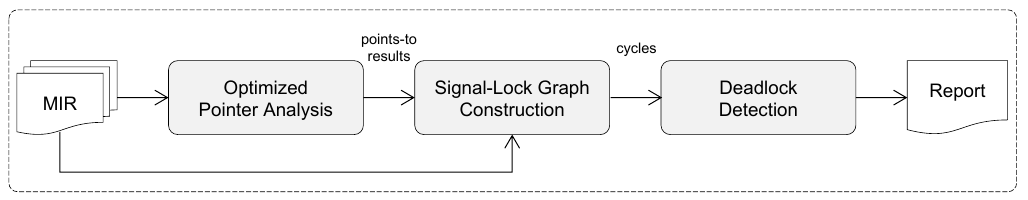}
  \caption{ The architecture of \textsc{RcChecker}}
  \label{framework}
\end{figure*}
In this section, we follow the architecture depicted in Figure \ref{framework} to detail the method and algorithms. 
The entire analysis process is organized into three distinct stages: 

\begin{enumerate}
    \item Optimized Pointer Analysis: This stage takes MIR as input, performs pointer analysis, and generates points-to results, which results serve as the foundation for constructing the signal-lock graph.
    \item Signal-Lock Graph Construction: This stage also takes MIR as input and combines it with pointer analysis results to construct the signal-lock graph. The cycles in the signal-lock graph serve as input for the next stage of deadlock detection.
    \item Deadlock Detection: This stage identifies cycles in the signal-lock graph to detect potential deadlocks and ultimately generates a detection report.
\end{enumerate}

\subsection{Optimized Pointer Analysis for Rust} \label{paS}
The first step in our static analysis approach involves conducting pointer analysis based on MIR, which establishes the groundwork for subsequent deadlock detection. We implement the Andersen-style pointer analysis \citep{andersen1994program} for Rust programs and propose two rules grounded in the ownership and smart pointer mechanisms to optimize the process. 

Drawing inspiration from SVF \citep{sui2016svf}, we adopt three loosely coupled components: constraint graph, rules, and solver. 
Each node in the constraint graph represents a variable $v$. The edges in the constraint graph represent the constraints between variables. 
Initially, we construct the constraint graph based on MIR statements, classifying them into specific patterns as detailed in Table \ref{opa}, each associated with a corresponding constraint-solving rule.
When the initial constraint graph is established, the solver component iteratively updates the points-to-sets ($pts$) of variables using a fixed-point algorithm guided by the rules defined in Table \ref{opa}.
\begin{table}[t]
\caption{Constraint generation rules for pointer analysis}
\label{opa}
\begin{tabular}{lc}
\toprule
\textsc{[Ref]} &  
$\dfrac{\gamma,\, s: p_{1} =\&p_{2} \quad v_{1}:(\gamma,p_{1}), \, v_{2}:(\gamma,p_{2})}
{v_{2} \in pts(v_{1})}$ \\[3ex] 

\textsc{[Copy]} &  
$\dfrac{\gamma,\, s: p_{1} =\texttt{(copy)}\,p_{2} \quad v_{1}:(\gamma,p_{1}), \, v_{2}:(\gamma,p_{2})}
{pts(v_{2}) \subseteq pts(v_{1})}$ \\[3ex]  

\textsc{[Move]} & 
$\dfrac{\gamma,\, s: p_{1} =\texttt{move}\;p_{2} \quad v_{1}:(\gamma,p_{1}), \, v_{2}:(\gamma,p_{2})}
{-}$ \\[3ex]  

\textsc{[Move-Rc]} & 
$\dfrac{\gamma,\, p_2:\texttt{Arc/Rc},\,s: p_{1} =\texttt{move}\;p_{2} \quad v_{1}:(\gamma,p_{1}), \, v_{2}:(\gamma,p_{2})}
{pts(v_{2}) \subseteq pts(v_{1})}$ \\[3ex]  

\textsc{[Load]} & 
$\dfrac{\gamma,\, s: p_{1} =*p_{2} \quad v_1:(\gamma,p_{1}), \, v_{2}:(\gamma,*p_{2}) \quad  v \in pts(v_2)}
{pts(v) \subseteq pts(v_{1})}$ \\[3ex]  

\textsc{[Store]} & 
$\dfrac{\gamma,\, s: *p_{1} =p_{2} \quad v_1:(\gamma,*p_{1}), \, v_{2}:(\gamma,p_{2}) \quad v \in pts(v_1)}
{pts(v_2) \subseteq pts(v)}$ \\[3ex]  

\textsc{[Field]} & 
$\dfrac{\gamma,\, s: p_{1} =p_{2}.x \quad v_1:(\gamma,p_{1}), \, v_{2}:(\gamma,p_{2}.x) \quad v \in pts(v_2)}
{pts(v_2) \subseteq pts(v_1)}$ \\[3ex]  

\textsc{[Call]}  &
$\dfrac{
  \begin{aligned}
  \gamma,\, s: p=\texttt{foo}(p_i), \,\gamma^{\prime}:\texttt{foo}([\&p_i],p_{ret})\{b_i\} \quad \\
  v:(\gamma,p_{i}), \, v^{\prime}:(\gamma^{\prime},p_{i}), \, v_{des}:(\gamma,p_{des}), \, v_{ret}:(\gamma^{\prime},p_{ret})
  \end{aligned}
}{
  pts(v) \subseteq pts(v^{\prime}), \, pts(v_{des}) \subseteq pts(v_{ret})
}$ \\[4ex] 

\textsc{[Call-Mv]}  &
$\dfrac{
  \begin{aligned}
  \gamma,\, s: p=\texttt{foo}(\texttt{move}\;p_i), \,\gamma^{\prime}:\texttt{foo}([p_i],p_{ret})\{b_i\} \\
  v:(\gamma,p_{i}), \, v^{\prime}:(\gamma^{\prime},p_{i}), \, v_{des}:(\gamma,p_{des}), \, v_{ret}:(\gamma,p_{ret})
  \end{aligned}
}{-}$ \\[4ex] 

\textsc{[Call-Rc]}  &
$\dfrac{
  \begin{aligned}
  \gamma,\, p_1:\texttt{Arc/Rc},\,s: p=\texttt{foo}(\texttt{move}\;p_i),\,\gamma^{\prime}:\texttt{foo}([p_i],p_{ret})\{b_i\} \\
  v:(\gamma,p_{i}), \, v^{\prime}:(\gamma^{\prime},p_{i}), \, v_{des}:(\gamma,p_{des}), \, v_{ret}:(\gamma,p_{ret})
  \end{aligned}
}{
  pts(v) \subseteq pts(v^{\prime}), \, pts(v_{des}) \subseteq pts(v_{ret})
}$ \\[4ex] 
\bottomrule
\end{tabular}
\end{table}

During assignments and function calls in Rust, ownership transfers frequently occur. In MIR, ownership transfer is indicated by the \verb|move| keyword. For most types, when ownership is transferred, such as in \verb|y| = \verb|move| \verb|x|, we can assume there is no longer a constraint relationship between \verb|y| and \verb|x|. This is because, at this point, the lifetime of \verb|x| has ended, and it will no longer be used. The Rust compiler also ensures no borrowing relationships involving \verb|x| at this stage. However, for reference-counted smart pointers \verb|Arc<T>| and \verb|Rc<T>|, the situation is different. Taking \texttt{Arc<T>} as an example, \verb|Arc<T>| are created from the \verb|T|, with the ownership of \verb|T| transferred to \texttt{Arc<T>}.  Subsequently, \texttt{Arc<T>} share and manage the ownership of \verb|T| through reference counting. When the reference count is not zero, even if ownership of \texttt{Arc<T>} is transferred, the data of type \verb|T| still exists as a single instance without generating any new copies. Therefore, we still preserve the corresponding constraint relationships when ownership transfers between reference-counted pointers. 

Specifically, in Table \ref{opa}, the patterns \textsc{move} and \textsc{call-mv} represent ownership transfer for ordinary variables, while \textsc{move-rc} and \textsc{call-rc} are dedicated to analyzing ownership transfer involving reference-counted smart pointers.

\begin{figure}[t] 
  \centering
  \includegraphics[width= \linewidth]{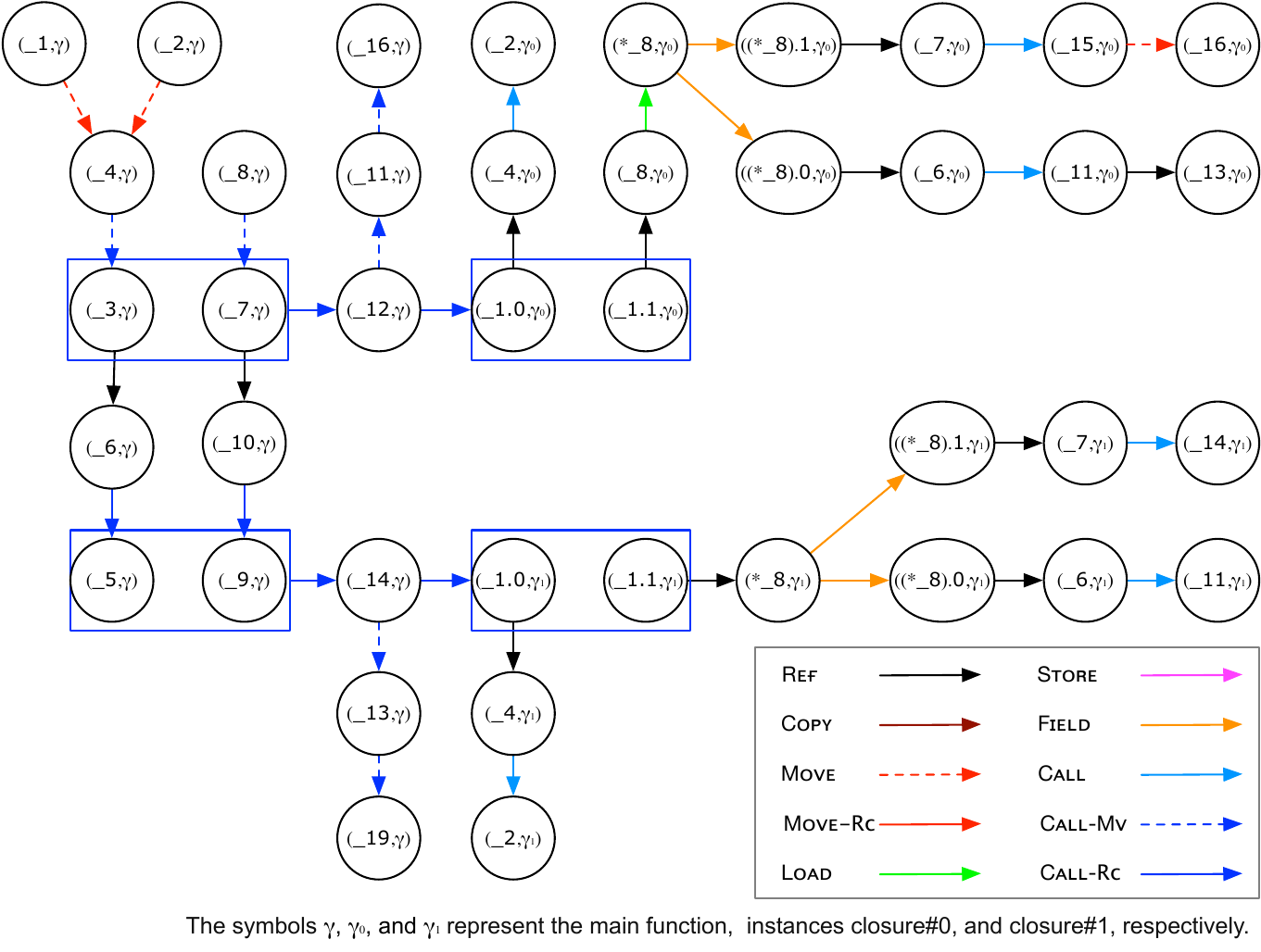}
  \caption{The initial constraint graph for pointer analysis corresponding to the example shown in Fig. \ref{bg}}
  \label{pa}
\end{figure}

\textbf{Example 4.1.} We use the MIR representation of the example in Figure \ref{bg} to illustrate this process. The symbols $\gamma$, $\gamma_0$, and $\gamma_1$ represent the main function, and the instances of \texttt{closure\#0} and \texttt{closure\#1}, respectively. Each place within a function instance corresponds to a node in the constraint graph. Based on each statement, edges are added between nodes, categorized into ten different types as defined in Table \ref{opa}. For ownership transfer cases, such as the statement \texttt{_4 = (move _1, move _2)} in \texttt{bb2} of the main function’s MIR, we introduce \textsc{Move} edges between \texttt{_1} and \texttt{_4}, as well as \texttt{_2} and \texttt{_4}. Similarly, for ownership transfer via function calls, such as \texttt{_3 = Arc::<(Mutex<bool>, Condvar)>::new(move _4)} in \texttt{bb2}, we add a \textsc{Call-Mv} edge between \texttt{_3} and \texttt{_4}. For closures, consider the statement \texttt{_12 = \{closure\text@7\}\{ mu1: move _7, pair1: move _3 \}}, where \texttt{_7} and \texttt{_3} are of type Arc. Since Arc is a reference-counted pointer, we establish \textsc{Call-Rc} edges between the function parameters, return values, and the arguments passed to the invoked function. The initial constraint graph for the code in Figure \ref{bg} is shown in Figure \ref{pa}. Once the initial constraint graph is constructed, we update the pts set gradually based on the rules in Table \ref{opa}.

\subsection{Signal-Lock Graph Construction} \label{slgS}

After completing the pointer analysis, we construct the signal-lock graph based on the lifetime information and the results of the pointer analysis. Building the signal-lock graph is divided into two steps: first, we collect signal-lock graph node information and maintain the set of alive nodes; second, we add signal-lock graph edges based on dependency and alias relationships. Subsequently, we elaborate on the two steps separately and use the example depicted in Figure \ref{lg} to illustrate the process.

\subsubsection{Creating Signal-Lock Graph Nodes}
\begin{algorithm}
\SetKwData{Left}{left}
\SetKwData{This}{this}
\SetKwData{Up}{up} 
\SetKwInOut{Input}{input}
\SetKwInOut{Output}{output}
	\Input{The MIR of a Rust program.} 
    \Output{$\mathcal{N}$, $\mathbb{A}$, $\mathbb{R}$}
	 \BlankLine 
        $CG \hookleftarrow$  construct a call graph for the MIR\;
        \ForEach{$\gamma$ $\in$ $CG.\Gamma$}{
            $\Sigma$ $\hookleftarrow$ the set of alive nodes;
            $ctx$ $\hookleftarrow$ the calling context; 
            $th$ $\hookleftarrow$ the thread\;
            \ForEach{$b$ $\in$ $\gamma.\mathcal{B}$}{
                $n \hookleftarrow$ create an empty node $\epsilon$ \;
                $n^{\prime} \hookleftarrow$ The node created from the predecessor basic block of $b$ \;
                 \If{$s:$ p = \textnormal{\texttt{Mutex::lock()}} $\in b.\mathcal{S}$}{
                   $n \hookleftarrow ((p,\gamma),\textsf{MX},l_s,ctx,th) $
                }
                \ElseIf{$s:$ p = \textnormal{\texttt{RwLock::read()}} $\in b.\mathcal{S}$}{
                 $n \hookleftarrow ((p,\gamma),\textsf{RL},l_s,ctx,th) $
                }
                 \ElseIf{$s:$ p = \textnormal{\texttt{RwLock::write()}} $\in b.\mathcal{S}$}{
                 $n \hookleftarrow ((p,\gamma),\textsf{WL},l_s,ctx,th) $
                }
                \ElseIf{$s:$ $p$ = \textnormal{\texttt{Condvar::wait(}}$p_{cv}$,$p_{mg}$\textnormal{\texttt{)}} $\in b.\mathcal{S}$}{
                 $n \hookleftarrow ((p_{cv},\gamma),\textsf{WT},l_s,ctx,th)$;
                 $\mathbb{R}(n) \hookleftarrow n_{\texttt{mg}}$\;
                }
                \ElseIf{$s:$ p =\textnormal{\texttt{ Condvar::notify(}}$p_{cv}$\textnormal{\texttt{)}} $\in b.\mathcal{S}$}{
                 $n \hookleftarrow ((p_{cv},\gamma),\textsf{NT},l_s,ctx,th)$\;
                }
               
                  \ElseIf{$s:p_{jh}=\textnormal{\texttt{spawn($p$)}}$}{
                   $ \gamma^{\prime} \hookleftarrow$ the thread closure $p$ corresponding instance;
                    $\gamma^{\prime}.\Sigma \hookleftarrow \Sigma$\;
                    $ctx^{\prime} \hookleftarrow$ the updated calling context; $\gamma^{\prime}.ctx \hookleftarrow ctx^{\prime}$\;
                    $th^{\prime} \hookleftarrow$ the new thread $(p_{jh},ctx^{\prime},\texttt{null})$
                } 
                 \ElseIf{$s:p= \textnormal{\texttt{join($p_{jh}$)}}$}{
                     $ctx^{\prime} \hookleftarrow$ the updated calling context\; 
                     $th \hookleftarrow$ corresponding to the $p_{jh}$; $th.ctx_{jn} \hookleftarrow ctx^{\prime}$
                } 
                 \ElseIf{$s:p=\textnormal{\texttt{func()}}$}{
               $ \gamma^{\prime} \hookleftarrow$ the instance generated by function call;
                    $\gamma^{\prime}.\Sigma \hookleftarrow \Sigma$\;
                    $ctx^{\prime} \hookleftarrow$ the updated calling context;$ \gamma^{\prime}.ctx \hookleftarrow ctx^{\prime}$\;
                }
                \ElseIf{$s:\textnormal{\texttt{drop(}}p\texttt{)}$ $or$ $s: p^\prime\textnormal{ = \texttt{move} p}\in b.\mathcal{S}$}{
                    $n_{p} \hookleftarrow$ the node corresponding to $v_\texttt{p}:(p,\gamma)$\;
                    Remove $n_{p}$ from $\Sigma$\;     
                }
                    \If{$n \neq \epsilon$}{
                    Add $n$ to $\mathcal{N}$\; 
                    $\mathbb{A}(n) \hookleftarrow \Sigma$;
                    Add $n$ to $\Sigma$\; }
                Add $n^{\prime}$ the $Pred(n)$ and $n$ the $Succ(n^{\prime})$\;
                
            }
        }
    \caption{Creating Signal-Lock Graph Nodes}
    \label{Algo Creating Signal-Lock Graph Nodes} 
 \end{algorithm}
The initial phase of constructing the signal-lock graph requires collecting node information and lifetime information, a process that is both flow-sensitive and context-sensitive. 
In Rust MIR, a basic block typically contains only one operation statement related to a lock or a condition variable, allowing us to map a node $n$ directly to a basic block $b$. 
We use $Succ(n)$ to record all nodes corresponding to the successor basic blocks of the basic block $b$ associated with $n$, and the set $Pred(n)$ contains all nodes corresponding to the predecessor basic blocks $b$ of the basic block associated with $n$; 
For every basic block, we create a corresponding node; however, a node only carries actual information if the basic block contains an operation related to a lock or condition variable. If the basic block does not contain such an operation, the node is considered empty, denoted as $n = \epsilon$ and primarily serves to preserve control-flow relationships, making it easier to identify loop structures in subsequent analyses.
The detailed process of creating signal-lock graph nodes is outlined in Algorithm \ref{Algo Creating Signal-Lock Graph Nodes}, and its core points can be summarized as follows.

\begin{itemize}
    \item \textit{Create signal-lock graph nodes}: 
    When traversing each basic block within each instance, we create nodes for scoped lock variables and condition variables. 
    When accessing a basic block, we first create an empty node $n=\epsilon$, where all variable information is initially set to null. Then, we traverse each statement in the basic block. If a statement is related to a lock or condition variable, we update the node information accordingly, as outlined in lines 6-15 of Algorithm \ref{Algo Creating Signal-Lock Graph Nodes}.
    
    \item \textit{Maintain and update \textnormal{$\mathbb{A}(n)$}}: We traverse each instance based on the call graph and maintain a set $\Sigma$ for each instance to record the currently alive non-empty nodes, initialized either as empty or set to the $\Sigma$ passed at the call site. 
    For each newly created node $n$, if $n$ is non-empty, we update $\mathbb{A}(n)$ to reflect the current $\Sigma$ and then add $n$ to $\Sigma$. 
    If a \texttt{drop(p)} or $p^\prime =$\texttt{move} $p$ operation occurs, we remove the node $n_\texttt{p}$ corresponding to variable $p$ from the current $\Sigma$. This ensures that we only record interactions between variables whose lifetimes overlap, thereby reducing false positives (e.g., the non-double lock scenario mentioned in Section 2.4).
    For function calls or thread creation, the $\Sigma$ of the called function or corresponding child thread instance is updated to mirror the current $\Sigma$. This process corresponds to lines 16-21 in Algorithm \ref{Algo Creating Signal-Lock Graph Nodes}.
\end{itemize}

Suppose there are $|\Gamma|$ instances in the MIR, each instance contains an average of $|\mathcal{B}|$ basic blocks, and each basic block consists of $|\mathcal{S}|$ statements. The time complexity of the algorithm is $O(|\Gamma|\times|\mathcal{B}|\times|\mathcal{S}|)$. The call graph $\Gamma$ is a finite set (the number of functions in the program is finite), and the basic blocks $\mathcal{B}$ in each function are also finite. Therefore, the entire algorithm can terminate in a finite number of steps.
After executing Algorithm \ref{Algo Creating Signal-Lock Graph Nodes}, we obtain $\mathbb{A}$, $\mathbb{R}$, and $\mathcal{N}$, which currently includes empty nodes that are removed later.

\begin{figure}[t] 
  \centering
  \includegraphics[width= \linewidth]{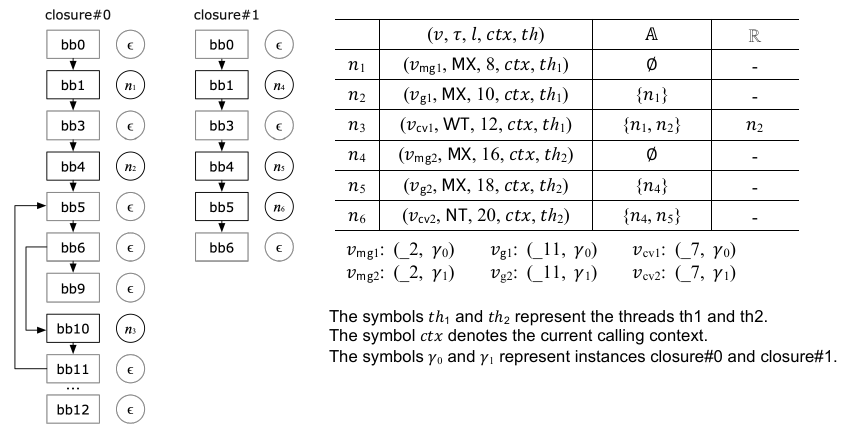}
  \caption{The process of creating signal-lock graph nodes for the example shown in Fig. \ref{bg}}
  \label{node}
\end{figure}

\textbf{Example 4.2.} 
We utilize the example depicted in Figure \ref{bg} to illustrate the process of creating nodes for the signal-lock graph. Let $ctx$ denote the calling context information, $th_1$ and $th_2$ represent \texttt{th1} and \texttt{th2}.
Figure \ref{node} illustrates the node creation process.
For the instance \texttt{closure\#0} $\gamma_0$ associated with \texttt{thread1}, we initialize \(\Sigma = \varnothing\) and then traverse the basic blocks.
First, we create an empty node for each basic block. When a basic block contains operations related to locks or condition variables, we update the node accordingly. For example, in MIR basic block \texttt{bb1}, the statement \texttt{_2 = Mutex::lock(copy 4)} (corresponding to line 8 in the source code) adds node information for  $v_\texttt{mg1}:(2,\gamma_0)$  as $(v_{\texttt{mg1}}, \textsf{MX}, 8, ctx, th_1)$, and we set $\mathbb{A}(n_1) = \varnothing$, then add $n_1$ to $\Sigma$. 
Similarly, in \texttt{bb4}, the statement \texttt{_11=Mutex::lock(copy _6)} (corresponding to line 10 in the source code) adds node information for  $v_\texttt{g1}:({11},\gamma_0)$  as $(v_{\texttt{g1}}, \textsf{MX}, 10, ctx, th_1)$. In \texttt{bb10}, the statement \texttt{_15=Condvar::wait(copy _7, move _17)} (corresponding to line 12 in the source code) adds node information for $v_\texttt{cv1}:(7,\gamma_0)$  as $v{\texttt{cv1}}, \textsf{WT}, 12, ctx, th_1$, and we set $\mathbb{A}(n_3) = \{n_1, n_2\}$. Additionally, based on the relationship between  $v_{\texttt{cv1}}$  and  $v_{\texttt{g1}}$ , we record $\mathbb{R}(n_3) = n_2$. The control flow relationships between basic blocks are recorded through the $Succ$ and $Pred$ sets of corresponding nodes, preserving loop structures.

For the instance \texttt{closure\#1} \texttt{$\gamma_1$} associated with thread \texttt{th2}, a similar process is followed. Nodes are created as follows: $n_4=(v_{\texttt{mg2}},\textsf{MX},16,ctx,th_2)$, $n_5=(v_{\texttt{g2}},\textsf{MX},18,ctx,th_2)$, $n_6=(v_{\texttt{cv2}}, \textsf{NT},20,ctx,th_2)$. And $\mathbb{A}(n)$ for each node is updated: $\mathbb{A}(n_4)=\varnothing$, $\mathbb{A}(n_5)=\{n_4\}$, $\mathbb{A}(n_6)=\{n_4,n_5\}$.

\subsubsection{Adding Signal-Lock Graph Edges}
After obtaining all the nodes and $\mathbb{A}(n)$, we start adding the edges of the signal-lock graph. We add dependency edges based on $\mathbb{A}(n)$ and alias edges based on the points-to information $pts$. The process, described in Algorithm \ref{Algo Adding Signal-Lock Graph Edges}, can be summarized in the following steps:
\begin{itemize}
    \item \textit{Eliminate empty nodes}: We check all empty nodes and iteratively remove them. If the predecessor node $n_i$ and successor node $n_j$ of an empty node $n$ are identical and non-empty, this indicates that the non-empty node $n_i$ resides within a loop. We add an edge $(n_i, n_j, \textsf{D})$ to represent the loop structure  (Lines 1-8).
    \item \textit{Add signal-lock graph edges}: The process is described in lines 8-16 of Algorithm \ref{Algo Adding Signal-Lock Graph Edges}. For a scoped lock node $n$ ($n.\tau \in \{\textsf{MX},\textsf{RL},\textsf{WL}\}$), for each $n_i \in \mathbb{A}(n)$, we add an edge $(n_i, n, \textsf{D})$ to signify that while holding lock $n_i$, lock $n$ is requested.
    
    We abstract \verb|Condvar::wait()| and \verb|Condvar::notify()| as operations for requesting and holding signal resources. When a thread waits for a signal to wake it up, we interpret the waiting action as requesting signal resources; when a thread executes a notify operation, we consider that the thread has been holding the signal resource until the \verb|notify()|. Therefore, for a condition variable node $n$, if $n.\tau=\textsf{WT}$ and $n_i \in \mathbb{A}(n)$, we add an edge $(n_i, n, \textsf{D})$ to indicate that the thread is requesting $n$ while holding $n_i$. If $n.\tau=\textsf{NT}$ and $n_i \in \mathbb{A}(n)$, we add an edge $(n, n_i, \textsf{D})$ to indicate that the thread has been holding $n$.
    
   Lastly, for nodes $n_i$ and $n_j$, we determine the aliasing relationship by comparing $pts(v_i)$ and $pts(v_j)$. If an aliasing relationship exists, we add an alias edge $(n_i, n_j, \textsf{A})$.
\end{itemize}

The time complexity is $O(|\mathcal{N}|^2)$, as it requires a double traversal of nodes to process edge relationships. Termination is guaranteed by the monotonically decreasing nature of the fixed node set $\mathcal{N}$ (removal of empty nodes) and the finite pointer set $pts$.
\begin{algorithm}[t]
\SetKwData{Left}{left}
\SetKwData{This}{this}
\SetKwData{Up}{up} 
\SetKwInOut{Input}{input}
\SetKwInOut{Output}{output}
	\Input{The set of signal-lock graph nodes $\mathcal{N}$, the points-to information $pts$.} 
	\Output{The signal-lock graph  $\mathcal{G}=(\mathcal{N},\mathcal{E})$.}
	 \BlankLine

        \ForEach{$n \in \mathcal{N}$}{
            \If{$n=\epsilon$}{
            
            \ForEach{$n_{i} \in Pred(n)$}{
               \ForEach{$n_{j} \in Succ(n)$}{
                \If{$n_{i}=n_{j}$ $\&$ $n_{i} \neq \epsilon$}{
                Add $(n_{i},n_{i},\textsf{D})$ to $\mathcal{E}$\;
                }
                Add $n_{j}$ to $Succ(n_{i})$, $n_{i}$ to $Pred(n_{j})$\;
            }
            }
            
            Remove $n$ from $\mathcal{N}, Succ(n_{i}),Pred(n_{j})$\;
            }
             
        }
        
        \ForEach{$n \in \mathcal{N}$}{
            \ForEach{$n_i\in \mathbb{A}(n)$}{
                    \If{$n.\tau$ $=$ $\textnormal{\textsf{NT}}$}{
                    Add $(n,n_i,\textsf{D})$ to $\mathcal{E}$\;
                    }\Else{
                    Add $(n_i,n,\textsf{D})$ to $\mathcal{E}$\;
                    }
            }
            \ForEach{$n_i\in \mathcal{N} \setminus{\{n\}}$}{
                \If{$pts(n_i.v) \subseteq pts(n.v)$ or $pts(n.v) \subseteq pts(n_i.v)$}{
                    Add $(n,n_i,\textsf{A})$ to $\mathcal{E}$\;
                }
            }
        }
    \caption{Adding Signal-Lock Graph Edges}
    \label{Algo Adding Signal-Lock Graph Edges} 
 \end{algorithm}

\begin{figure}[t] 
  \centering
  \includegraphics[width= \linewidth]{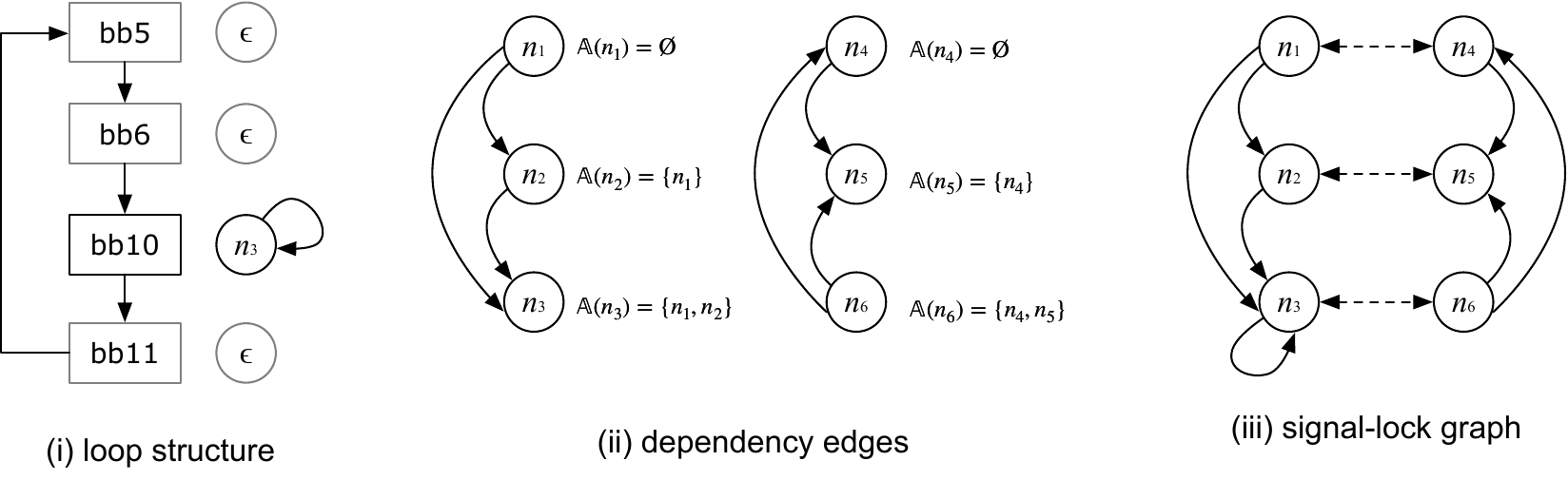}
  \caption{The process of adding signal-lock graph edges for the example shown in Fig. \ref{bg}}
  \label{edge}
\end{figure}

\textbf{Example 4.3.} 
We preserve control flow relationships by creating empty nodes for basic blocks. Taking the example in Figure \ref{bg}, for \texttt{closure\#0}, the control flow relationship $bb5\rightarrow bb6\rightarrow bb10\rightarrow bb11\rightarrow bb5$ is recorded through $Succ$ and $Pred$ to represent the control flow between corresponding nodes, as shown in Figure \ref{edge}(i). We iteratively eliminate empty nodes, and when $Succ(n_3)=n_3$ and $Pred(n_3)=n_3$, we identify this loop structure. Then, we add a dependency edge $(n_3, n_3, \textsf{D})$ in the signal lock graph.

For scoped lock variable nodes, we add dependency edges based on $\mathbb{A}(n)$, resulting in $(n_1, n_2, \textsf{D})$ and $(n_4, n_5, \textsf{D})$. For the wait operation, we add dependency edges $(n_1, n_3, \textsf{D})$ and $(n_2, n_3, \textsf{D})$. And for the notify operation, we add dependency edges $(n_6, n_4, \textsf{D})$ and $(n_6, n_5, \textsf{D})$. Finally, using points-to information, we add edges $(n_1, n_4, \textsf{A})$, $(n_2, n_5, \textsf{A})$, $(n_3, n_6, \textsf{A})$.

\subsection{Resource and Communication Deadlock Detection} \label{ddS}

After constructing the signal-lock graph, we traverse it to identify cycles that contain at least one dependency edge, which we define as dependency cycles $(\mathcal{C}\in \mathbb{C})$. 
Dependency cycles represent potential deadlocks. Based on the threads in which the nodes of the dependency cycle are located, we categorize dependency cycles into intra-thread dependency cycles and inter-thread dependency cycles.

First, we define a set of operations to describe the properties of the set $\mathcal{N}$ and $\mathcal{E}$. 
We use the method $th(\mathcal{N})$ retrieve the threads associated with all nodes in $\mathcal{N}$, defined as $th(\mathcal{N}) = \{ n.th \mid n \in \mathcal{N} \}$.
The method $\tau(\mathcal{N})$ retrieves the types of all nodes in $\mathcal{N}$, given by $\tau(\mathcal{N}) = \{ n.\tau \mid n \in \mathcal{N} \}$.
The method $D(\mathcal{E})$ filters out all dependency edges in $\mathcal{E}$, formally defined as $D(\mathcal{E}) = \{ e \in \mathcal{E} \mid e.k = \textsf{D} \}$.

We classify intra-thread dependency cycles into double-lock cycles $\mathbb{C}_{dl}$ and single-node cycles $\mathbb{C}_{sn}$.
Equations (1) and (2) provide the definitions.
A double lock cycle that contains two nodes $\mathcal{C} \in \mathbb{C}_{dl}$, indicating a potential double-lock structure, have a dependency edge and an alias edge (need to form a cyclic path).
A signal node cycle that contains only one node $\mathcal{C} \in \mathbb{C}_{sn}$ indicates a loop structure, with only one dependency edge. 
\begin{equation}
  \mathbb{C}_{dl}=\{\mathcal{C}|\mathcal{C} \in \mathbb{C}\land|th(\mathcal{N}_c)|=1\land |\mathcal{N}_c|=2\}
\end{equation}
\begin{equation}
  \mathbb{C}_{sn}=\{\mathcal{C}|\mathcal{C} \in \mathbb{C}\land|th(\mathcal{N}_c)|=1\land |\mathcal{N}_c|=1\}
\end{equation}

We define cycles that arise from conflicts in the order of inter-thread resource access as conflict cycles, denoted as $\mathbb{C}_c$.
Equation (3) provides the definition. For cycle $\mathcal{C} \in \mathbb{C}_c$, there are dependency edges between nodes within the same thread and alias edges between different threads, forming a complete cyclic path. 
\begin{equation}
  \mathbb{C}_c=\{\mathcal{C}|\mathcal{C} \in \mathbb{C}\land|th(\mathcal{N}_c)|>1\land |D(\mathcal{E}_c)|=|th(\mathcal{N}_c)|\}
\end{equation}

Based on the type of nodes involved, conflict cycles $\mathbb{C}_c$ can be categorized into two types: conflict lock cycles $\mathbb{C}_{cl}$ and conflict signal-lock cycles $\mathbb{C}_{csl}$. 
Equations (4) and (5) provide the definitions. A cycle $\mathcal{C} \in \mathbb{C}_{cl}$ includes only scoped lock variable nodes, which indicate a potential conflict lock structure. In contrast, a cycle $\mathcal{C} \in \mathbb{C}_{csl}$ includes condition variable nodes, which represent a potential conflict signal-lock structure.
\begin{equation}
  \mathbb{C}_{cl}=\{\mathcal{C}|\mathcal{C} \in \mathbb{C}_c\land \textsf{WT}\notin \tau(\mathcal{N}_c) \land \textsf{NT}\notin \tau(\mathcal{N}_c)\}
\end{equation}
\begin{equation}
  \mathbb{C}_{csl}=\{\mathcal{C}|\mathcal{C} \in \mathbb{C}_c \land (\textsf{WT}\in \tau(\mathcal{N}_c) \lor \textsf{NT}\in \tau(\mathcal{N}_c))\}
\end{equation}

\IncMargin{1em}
\begin{algorithm}[t]
\SetAlgoNlRelativeSize{0}
\SetNlSty{textbf}{}{}
\SetAlgoNlRelativeSize{-1}
\SetKwData{Left}{left}\SetKwData{This}{this}\SetKwData{Up}{up}
\SetKwInOut{Input}{input}\SetKwInOut{Output}{output}

\Input{The signal-lock graph $\mathcal{G}=(\mathcal{N},\mathcal{E})$.}
\Output{Deadlock detection report $\mathcal{DS}$.}
\BlankLine
\SetKwFunction{MyFunction}{CheckLoop} 
\SetKwProg{Fn}{Function}{:}{\KwRet} 
$\mathcal{DS} = \varnothing$\;
$\mathbb{C}_{dl}, \mathbb{C}_{sn}, \mathbb{C}_{cl}, \mathbb{C}_{csl}$ $\hookleftarrow$ traverse $\mathcal{G}$ to update the dependency cycle sets\;
\Fn{\MyFunction{$n$}}{
    \ForEach{$\mathcal{C}=(\mathcal{N}_c,\mathcal{E}_c) \in \mathbb{C}_{sn}$}{
        \If{$n \in \mathcal{N}_c$}{
            \KwRet{$true$}\;
        }
    }
    \KwRet{$false$}\;
}

\ForEach{$\mathcal{C}=(\mathcal{N}_c,\mathcal{E}_c) \in \mathbb{C}_{dl}$}{
        $n_1$,$n_2$ $\hookleftarrow$ two nodes in $\mathcal{N}_c$\;
        \If{\textnormal{(}$n_1.\tau$,$n_2.\tau$\textnormal{)} $\in$ $\textnormal{\{(\textsf{MX}, \textsf{MX}), (\textsf{RL}, \textsf{WL}), (\textsf{WL}, \textsf{WL}), (\textsf{WL}, \textsf{RL})\}}$}{
            Add $(\mathcal{N}_c,\mathcal{E}_c,\textit{"double lock"})$ to $\mathcal{DS}$ \;
        }
    }
\ForEach{$\mathcal{C}=(\mathcal{N}_c,\mathcal{E}_c) \in \mathbb{C}_{cl}$}{
    $(n_1,n_2,\textsf{D})$, $(n_3,n_4,\textsf{D})$ $\hookleftarrow$ two dependency edges in $\mathcal{E}_c$\;
  \If{$!non\_concurrency(\mathcal{C})$}{
                Add $(\mathcal{N}_c,\mathcal{E}_c,\textit{"conflict lock"})$ to $\mathcal{DS}$ \;
    }
}
\ForEach{$\mathcal{C}=(\mathcal{N}_c,\mathcal{E}_c) \in \mathbb{C}_{csl}$}{
       $(n_1,n_2,\textsf{D})$, $(n_3,n_4,\textsf{D})$ $\hookleftarrow$ two dependency edges in $\mathcal{E}_c$\;
            \If{$\mathbb{R}(n_{2}) \neq n_1$ $\&$ $\mathbb{R}(n_{4}) \neq n_3$ $\&$$!non\_concurrency(\mathcal{C})$}{
                Add $(\mathcal{N}_c,\mathcal{E}_c,\textit{"conflict signal-lock"})$ to $\mathcal{DS}$ \;
            }
       
    }
$\mathcal{N}_{\textsf{wt}}$ $\hookleftarrow$  all \textsf{WT} type nodes in $\mathcal{N}$\;
\ForEach{$n_{\textnormal{\textsf{wt}}} \in \mathcal{N}_{\textnormal{\textsf{wt}}}$}{
        $e:(n_{\textsf{wt}},n_{\textsf{nt}},\textsf{A})$ $\hookleftarrow$ the alias edge connected to $n_{\textsf{wt}}$\;
        $N_{\textsf{lock}}$ $\hookleftarrow$ the node $\mathbb{R}(n_{\textnormal{\textsf{wt}}})$ and its alias nodes\;
            \If{$ N_{\textnormal{\textsf{lock}}} \cap \mathbb{A}(n_{\textnormal{\textsf{nt}}}) \neq \varnothing$ $\&$ \MyFunction{$n_{\textnormal{\textsf{wt}}}$}}{
            continue\;
            }
            Add $(\{n_{\textsf{wt}},n_{\textsf{nt}}\},\{e\},\textit{"lost notification"})$ to $\mathcal{DS}$\;      
        % }
         
}

\caption{Resource and Communication Deadlock Detection}
\label{Detection for deadlock}
\end{algorithm}
\DecMargin{1em}

To address the potential false positives caused by impossible concurrent execution, we introduce the function $non\_concurrency(\mathcal{C})$ to determine whether operations within a cycle can execute concurrently. This function performs the following judgments:
First, for determining whether threads can execute concurrently, we utilize the recorded thread triple information $(p_{jh}, ctx_{sp}, ctx_{jn})$ for preliminary analysis. For two threads, if the \texttt{join()} operation of thread 1 occurs before the \texttt{spawn()} operation of thread 2, these two threads cannot execute concurrently. Although this mechanism does not handle complex scenarios involving cyclic thread creation without explicit \texttt{join()} operations, it effectively filters out some simple false positives.
Additionally, we detect the presence of a gatelock mechanism. A gatelock \citep{havelund2000using} refers to a situation where, in two threads with conflicting lock acquisition orders, if they synchronize using the same lock (or aliasing locks) before entering the conflicting lock region, the conflicting lock acquisition does not lead to a deadlock and should be considered a non-concurrent region. We further extend the concept of gatelock to condition variables, meaning that resource access conflicts not only include conflicting lock patterns but also conflicting signal lock patterns. When the variable acting as the “gate” is a condition variable, it may manifest as follows: before conflicting access occurs, a thread synchronizes via \texttt{wait()}, and only after another thread executes \texttt{notify()} does the first thread proceed. If such a protective mechanism exists, the related threads are not considered capable of concurrent execution.

We traverse the signal-lock graph to identify cycles that correspond to the specified patterns and subsequently conduct deadlock detection. The algorithm is detailed in Algorithm \ref{Detection for deadlock} and comprises the following four steps:
\begin{itemize}
    \item \textit{double lock detection}:  If $\mathcal{C}\in \mathbb{C}_{dl}$ and the two nodes in $\mathcal{C}$ have types that match one of the following pairs: $(\textsf{MX}, \textsf{MX})$, $(\textsf{RL}, \textsf{WL})$, $(\textsf{WL}, \textsf{WL})$, or $(\textsf{WL}, \textsf{RL})$, then it represents a double-lock structure (Lines 8-11). 
    \item \textit{conflict lock detection}:  If $\mathcal{C}\in \mathbb{C}_{cl}$, we need to assess the concurrency relationships of the threads and determine if there is any gatelock protection, which is represented by $non\_concurrency()$ in Algorithm \ref{Detection for deadlock}. If threads can execute concurrently, we report it as a conflict lock (Lines 12-15).
    \item \textit{conflict signal-lock detection}:  If $\mathcal{C}\in \mathbb{C}_{csl}$, we need to assess further the concurrency relationships between the threads and $\mathbb{R}$. If the threads can execute concurrently and the dependency cycle $\mathcal{C}$ does not occur between the $\textsf{WT}$ node of the condition variable and the associated lock, we report it as a conflict-signal lock (Lines 16-19).
    \item \textit{Lost notification detection}: We evaluate the $\textsf{WT}$ and $\textsf{NT}$ nodes in $\mathcal{N}$. If the $\textsf{WT}$ node is also present in another cycle $c^{\prime} \in \mathbb{C}_{sn}$, we consider it to conform to the proper use of wait (lines 3-7). If it is not simultaneously present in such a cycle, the wait operation is typically a standalone wait or in a branching structure, which poses the risk of missed notifications and spurious wake-ups. If the node corresponding to the lock associated with the condition variable is not present in the $\mathbb{A}(n)$ set of the $\textsf{NT}$ node, we assume that the thread containing the notify operation has not updated the condition, which poses a risk of permanent blocking. Therefore, we uniformly report such cases as lost notifications (Lines 20-26).
\end{itemize}

The time complexity is $O(\mathcal{|N|}\times\mathcal{|C|}|)$, where $\mathcal{|C|}$ is the number of dependency cycles and $\mathcal{|N|}$ is the total number of nodes. Termination is ensured by the finiteness of the fixed input graph G and the operations performed.
 
\begin{figure}[t] 
  \centering
    \includegraphics[width= 0.9\linewidth]{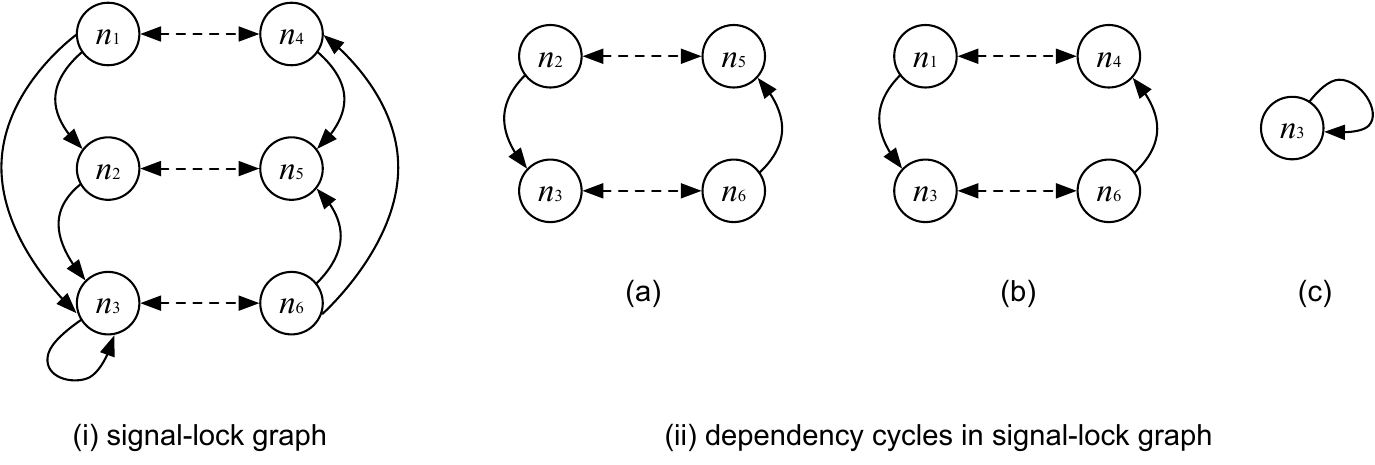}
    \caption{ Signal-lock graph cycles for the example in Fig. \ref{bg}}
    \label{cycle}
\end{figure}

\textbf{Example 4.4.} Traverse the signal-lock graph in Figure \ref{edge} to identify the dependency cycles, as depicted in Figure \ref{cycle}. First, both cycles (a) and (b) belong to $\mathbb{C}_{csl}$. We then check whether the threads containing the nodes within the dependency cycle can execute concurrently.  Here, $th_1$ and $th_2$ can be executed concurrently. Next, we examine the $\mathbb{R}$, where $\mathbb{R}(n_3) = n_2$. Therefore, cycle (a) occurs between the condition variable and its associated lock, leading us to exclude (a) and report (b) as a conflict signal-lock error. 

Next, we examine the \textsf{WT}-type nodes. If a \textsf{WT} node appears in a $\mathbb{C}_{cl}$ cycle (c), it indicates that the wait operation is used correctly. Additionally, the set of alive nodes for the \textsf{NT} node includes the lock node (or its alias node) associated with the condition variable, i.e., $n_5 \in \mathbb{A}(n_6)$. This suggests that the notify operation is also used correctly, allowing us to conclude that there is no lost notification error. Therefore, the code shown in Figure \ref{lg} contains only one conflict signal-lock error, which occurs between \verb|mutex| and \verb|condvar|, involving the relevant operations at lines 8, 12, 16, and 20.

It is noted that inferring whether the condition was correctly updated solely based on whether the associated lock is acquired before the notify operation is not sufficiently precise. For example, if the code at line 19 is commented out, the wait operation at line 12 will re-enter the wait state after being awakened. However, since the lock at line 18 was acquired, we might incorrectly assume that the notify thread executed the operation correctly, resulting in a false negative. Detecting such error patterns requires adding path-sensitive analysis, which is part of our future work.

\section{Evaluation} \label{section4}
\textsc{RcChecker} is developed using rustc version nightly-2023-09-13. The development and testing environments are both set up on a machine running Ubuntu 23.10, equipped with a 16-core 12th Gen Intel(R) Core(TM) i9-12900K CPU and 32GB of physical memory.

To evaluate \textsc{RcChecker}, we consider the following questions:
\begin{enumerate}
\item[\textbf{Q1}] How efficient and effective is \textsc{RcChecker} compared to the previous Rust static resource deadlock detection tool? What advantages does \textsc{RcChecker} offer? (Section \ref{Comparing with Other Tool})
\item[\textbf{Q2}] How does \textsc{RcChecker} perform in detecting resource and communication deadlocks simultaneously? Is it capable of detecting new unreported deadlocks? (Section \ref{section Applying to Real Programs})
\end{enumerate}

\subsection{Q1: Comparison with Existing Tool} \label{Comparing with Other Tool}
We compare \textsc{RcChecker} with the most recent and relevant open-source tool for Rust deadlock detection, \textsc{Lockbud} \citep{Boqin2020understanding,lockbud,qin2024understanding}. \textsc{Lockbud} is a static analysis tool that detects concurrency and memory errors. 
The deadlock detection module of \textsc{Lockbud} primarily targets resource deadlocks. \textsc{Lockbud} performs intra-procedural pointer analysis and uses type information to infer inter-procedural alias relationships.

We follow the benchmarks from \textsc{Lockbud} and conduct comparative experiments on four small-scale programs and eighteen real-world applications. The four small-scale programs are collected and adapted from \textsc{Lockbud}’s built-in test cases. 
For the eighteen real-world applications, we collect 79 real deadlocks from \textsc{Lockbud}, all of which are resource deadlocks. Table \ref{compare with lockbud} provides detailed information on the benchmarks. 

The overhead in Table \ref{compare with lockbud} records the overhead of \textsc{Lockbud} and \textsc{RcChecker}, calculated as the ratio of detection time to the project build time. 
The average overhead for \textsc{RcChecker} is 15.47\%, compared to 7.16\% for \textsc{Lockbud}. \textsc{RcChecker} incurs 8.31\% more overhead than \textsc{Lockbud} due to its ability to detect a wider range of deadlocks and its improved precision.
Moreover, the overhead introduced by \textsc{RcChecker} is within an acceptable range relative to the project build time.

\begin{figure}[t] 
  \centering
  \includegraphics[width= \linewidth]{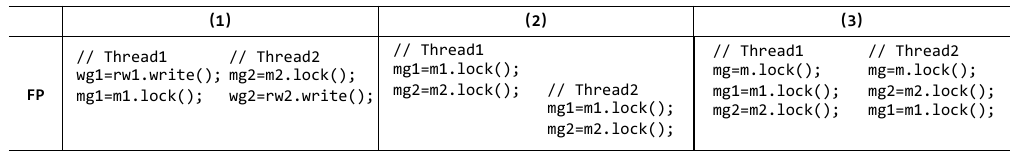}
  \caption{False positives in \textsc{Lockbud}}
  \label{fp-fn}
\end{figure}

The report in Table \ref{compare with lockbud} records the number of deadlocks reported by \textsc{RcChecker} and \textsc{Lockbud}, with \textbf{FP} indicating false positives. \textsc{RcChecker} identifies eight false positives in \textsc{Lockbud}. The eight false positives are verified through our manual inspection, and we provide a detailed summary in Figure \ref{fp-fn}, categorizing them into three types of false positives. The first type of false positive occurs because \textsc{Lockbud} uses type inference to determine inter-procedural aliasing relationships, leading to the misidentification of different locks within separate threads as being the same due to their identical types. The second false positive stems from ignoring the concurrent relationships between threads. The third false positive arises from overlooking the protection of gatelocks. \textsc{RcChecker} eliminates these false positive scenarios by employing the optimized pointer analysis, enhancing detection precision.

\begin{table}[t]
\caption{Comparison of experimental results}
\label{compare with lockbud}
\setlength{\tabcolsep}{1pt}
\begin{tabular*}{\textwidth}{@{\extracolsep\fill}lccccccc}
\toprule%
& \multicolumn{3}{@{}c@{}}{\textbf{Application Info\footnotemark[1]}} & \multicolumn{2}{@{}c@{}}{\textsc{\textbf{RcChecker}}}  &\multicolumn{2}{@{}c@{}}{\textsc{\textbf{Lockbud}}}\\
\cmidrule{2-4}\cmidrule{5-6} \cmidrule{7-8}
& KLOC & Count & Type& Overhead\footnotemark[2] & Deadlock &Overhead\footnotemark[2] & Deadlock\footnotemark[3] \\
\midrule
type-same&1&   0& -  &5.43\%&0&4.31\%&2 (\textbf{2 FP}) \\
gatelock&1&   0& - &5.64\%&0& 3.32\%& 1 (\textbf{1 FP}) \\
thread-join&  0& 0& - &8.79\%&0&6.79\% & 1 (\textbf{1 FP})  \\
% \cmidrule{1-11} 
Redox(tfs) & 11   &0&- &19.37\% & 0 & 6.85\%  & 0\\
RCore & 22 &   8 & I, II   & 25.45\% & 8 & 7.79\%  & 8 \\
Rayon & 24 &  0&- & 15.55\% & 0& 5.52\%  & 0 \\
Crossbeam & 26 & 0&- & 17.64\%& 0 & 6.39\%  & 0 \\
Grin & 44 &  3 &I, II & 25.71\%& 3 & 7.86\%  & 3 \\
Deno & 56 &  0&-& 25.23\%  & 0 & 8.80\%& 0 \\
Firecracker & 61 &  0&- & 23.51\% & 0  & 9.81\% & 0 \\
Lighthouse & 102 &  9 &I& 25.59\% & 9 & 14.82\% & 9 \\
Tock & 117 &  0 &-&  15.66\% & 0 & 4.42\% & 0 \\
Openethereum & 128 & 27&I, II& 14.44\%& 27 & 7.27\%  & 30 (\textbf{3 FP}) \\
Solana & 214 & 17&I, II& 5.34\%& 17 & 3.87\%  & 17 \\
TiKV & 237 & 0  &-& 10.90\%& 0 & 8.90\%  & 0 \\
Substrate & 247  &2&I & 4.11\% & 2& 4.92\%  & 2\\
Diem & 258 & 0  &-& 4.36\%& 0 & 5.12\%  & 0 \\
Winit & 291 & 1&I& 15.78\% & 1& 7.41\%  & 1 \\
Serenity & 301 &0 &- & 23.95\% & 0& 10.21\%  & 0 \\
Servo & 320  &0&- & 9.94\% & 0& 5.92\%  & 1 (\textbf{1 FP}) \\
Wasmer & 751  &12&I & 22.67\% & 12& 10.24\% & 12 \\
\botrule
\end{tabular*}
\footnotetext[1]{Records the number and types of deadlocks found in the program, with types I and II corresponding to double lock and conflict lock deadlock types.}
\footnotetext[2]{Indicates the ratio of detection time to project build time.}
\footnotetext[3]{Records the number of detected deadlocks, where \textbf{FP} represents false positives.}

\end{table}

\begin{tcolorbox}[colback=white, sharp corners, boxrule=0.5pt] 
Answer to Q1: Compared to the state-of-the-art approach, \textsc{RcChecker} demonstrates better precision in detecting resource deadlocks, effectively reducing three types of false positives. 
\end{tcolorbox}

\subsection{Q2: New Deadlock Findings} \label{section Applying to Real Programs}

To evaluate the capability of \textsc{RcChecker} in detecting communication deadlocks and to further explore its advantages in identifying resource deadlocks, we conduct tests on 11 programs.

Table \ref{Applying to Real Programs.} lists relevant information about the 11 applications, including the program size and the number of \verb|Mutex|, \verb|RwLock|, and \verb|Condvar|.
The column of \textbf{SLG Info} in Table \ref{Applying to Real Programs.} represents the number of nodes and edges recorded in the signal-lock graph.
The column of \textbf{Time} records the build time of the programs, the detection time of \textsc{RcChecker}, and overhead(i.e., the ratio of detection time to build time of \textsc{RcChecker}). 
The detection time of \textsc{RcChecker} encompasses the duration of pointer analysis, the construction of SLG, and the deadlock detection process based on the SLG.
The overhead of \textsc{RcChecker} ranges from 3\% to 26\%, which indicates that the time overhead introduced by \textsc{RcChecker} is acceptable compared to the project build time. 
The column of \textbf{Report} records the number and types of deadlocks reported by \textsc{RcChecker}. 
In the tested programs, \textsc{RcChecker} detects a total of 17 deadlocks, with 7 originating from real-world applications. 
Exiting tool \textsc{Lockbud} fails to detect these errors.
The first five are test cases we designed ourselves, containing nine communication deadlocks and one type of resource deadlock error that \textsc{Lockbud} fails to detect.
The remaining six test cases are from real-world systems, covering a range of applications including blockchain, web systems, concurrent frameworks, and embedded systems \citep{log4rs,accesskit,burble,kaskada,conflux}.
In these six applications, \textsc{RcChecker} report seven new deadlocks, including two resource deadlocks (both being double lock errors) and five communication deadlocks (one conflict signal-lock error and four lost notification errors).

\textsc{Lockbud} does not support the detection of communication deadlocks and fails to detect a type of resource deadlock, namely the case we demonstrated in Section \ref{typeS} on double lock, which arises from the lack of inter-procedural pointer analysis. When locks are passed as function parameters, \textsc{Lockbud} fails to resolve alias relationships.

\begin{table}[t]
\caption{New deadlock findings}
\label{Applying to Real Programs.}
\setlength{\tabcolsep}{2pt}
\begin{tabular*}{\textwidth}{@{\extracolsep\fill}lcccccccc}
\toprule%
& \multicolumn{1}{@{}c@{}}{}{}
& \multicolumn{2}{@{}c@{}}{}{\textbf{SLG Info\footnotemark[1]}}
&\multicolumn{3}{@{}c@{}}{\textbf{Time\footnotemark[2]}} 
& \multicolumn{2}{@{}c@{}}{\textbf{Report\footnotemark[3]}} \\
\cmidrule{3-4}\cmidrule{5-7}\cmidrule{8-9}%

 & KLOC&Node & Edge  &Build&Detection&Overhead& Deadlock & Type\\
\midrule
cvar-closure&0.3&  16&  14 & 0.96s&0.03s&3.13\%&2 &III, IV\\
cvar-struct&0.3&  15&  11 & 0.64s&0.02s&3.14\%&2 &III, IV\\
notify-lost &1& 4&  3 & 1.17s&0.05s&4.27\%&2 &IV\\
improper-cvar&1& 4& 5& 1.66s&0.07s&4.21\%&3 &III\\
swap-deadlock&1& 2& 2& 1.29s&0.05s&3.87\%&1 &I\\

Arcaders & 2 & 6&   10& 3.2s&0.1s&3.12\%&1 &III\\
Log4rs & 7 & 15 &   13 & 14s&3.4s&24.28\%&1  & IV  \\
AccessKit & 19 & 33 & 21   & 25s&6.2s&24.80\%&1 & IV  \\
Burble & 26 & 21 & 14 &  32s&7.3s&22.81\%&2 & I  \\
Kaskada & 94 & 46 & 21  & 142s&36s&25.35\%&1 & IV  \\
Conflux-Rust & 323 & 672& 320 &2746s &389s&14.10\%&1 & IV  \\
\botrule
\end{tabular*}
\footnotetext[1]{Records the number of nodes and edges recorded in the signal-lock graph.}
\footnotetext[2]{Record the project build time and RcChecker detection time, with overhead expressed as the ratio of detection time to project build time.}
\footnotetext[3]{Documents the number of reported deadlocks and the types involved, with types I, III, and IV corresponding to double lock, conflict signal-lock, and lost notification deadlock types.}
\end{table}

\begin{figure}[t] 
  \centering
    \includegraphics[width=\linewidth]{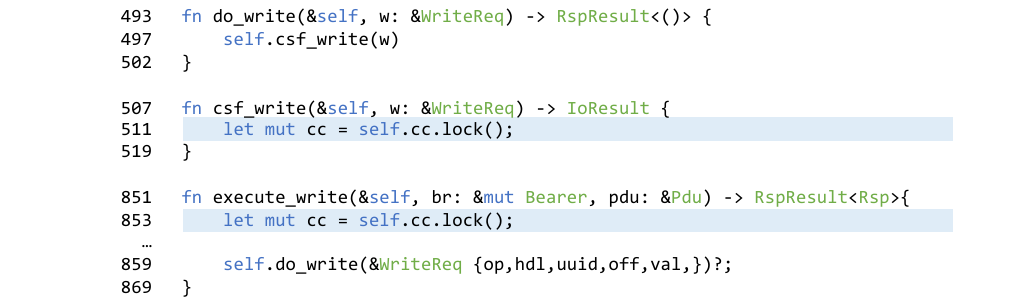}
    \caption{A resource deadlock in Burble reported by \textsc{RcChecker}}
    \label{real1}
\end{figure}
\begin{figure}[t] 
  \centering
    \includegraphics[width=\linewidth]{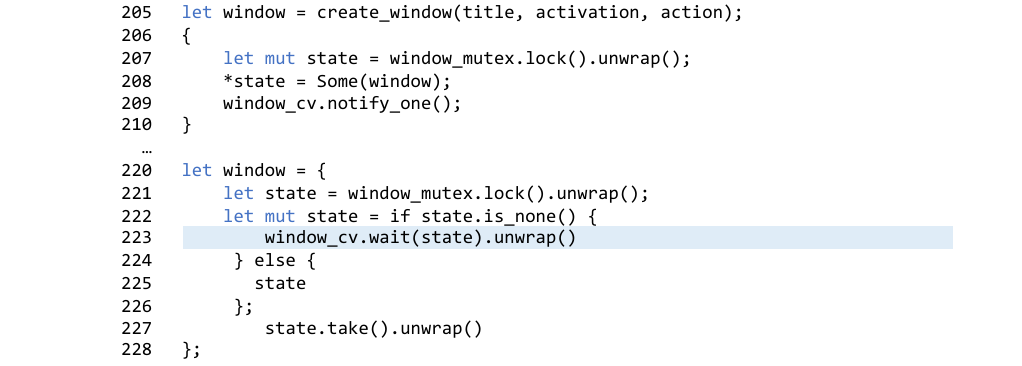}
    \caption{A communication deadlock in AccessKit reported by \textsc{RcChecker}}
    \label{real2}
\end{figure}

Figure \ref{real1} depicts a double lock reported in Burble (\citeyear{burble}). In line 853, a \verb|MutexGuard| variable \verb|cc| is returned by calling \verb|self.cc.lock()|, and its lifetime extends until line 869, where the \verb|execute_write()| function ends. During the function call in line 859, the \verb|do_write()| function in line 497 is invoked, eventually leading to the execution of \verb|self.cc.lock()| at line 511. Since the \verb|cc| variable created in line 853 is still within its lifetime, a double lock occurs. Figure \ref{real2} depicts a lost notification issue reported in AccessKit (\citeyear{accesskit}). The \verb|window_cv.wait(state)| at line 223 is not enclosed within a loop structure, which can result in the notify operation being executed before the wait operation and spurious wakeups as discussed in Section \ref{tsS}. Thus, \textsc{RcChecker} reports it as a lost notification.

\begin{tcolorbox}[colback=white, sharp corners, boxrule=0.5pt]
Answer to Q2: \textsc{RcChecker} effectively detects communication deadlocks and addresses a previously existing false negative issue in the existing tool. \textsc{RcChecker} newly reports seven previously unreported deadlocks in real-world systems, including two resource deadlocks and five communication deadlocks.
\end{tcolorbox}

\section{Limitations} \label{section5}
\textsc{RcChecker} has some limitations: 1) \textsc{RcChecker} cannot precisely detect communication deadlocks caused by condition predicates. We plan to make \textsc{RcChecker} path-sensitive in the future to achieve better precision. 2) \textsc{RcChecker} currently supports deadlock detection for synchronization mechanisms based on locks and condition variables. We plan to extend its capabilities further to support additional Rust synchronization primitives, such as \texttt{Channel} and \texttt{Semaphore}. At the same time, we are also considering handling the asynchronous features of Rust.

\section{Related Work} \label{section6}
\subsection{Deadlock Detection}
Deadlock detection is an essential task for ensuring the reliability of concurrent programs and is divided into resource deadlock detection and communication deadlock detection.

Resource deadlock detection has been extensively studied in the past, with the main approaches being dynamic and static analysis methods. Dynamic analysis \citep{bensalem2005dynamic, joshi2009randomized, cai2014magiclock, cai2020low, zhou2022deadlock,cai2012magicfuzzer,eslamimehr2014sherlock,jula2008deadlock,samak2014trace,sorrentino2015picklock,zhou2017undead} identifies potential deadlocks by monitoring a program’s behavior at runtime, but its coverage is limited. In contrast, static analysis inspects a program’s source code to detect potential deadlocks, effectively analyzing obscure program paths. \textsc{RacerX} \citep{engler2003racerx} targets C programs and uses flow-sensitive interprocedural analysis to detect race conditions and deadlocks. It caches lockset results at the function level to achieve context sensitivity, which may lead to scalability issues in large systems. Williams et al. (\citeyear{williams2005static}) construct a global context-sensitive lock graph for Java libraries. Their tool generally reports false positives due to non-concurrency and path-infeasible deadlocks. \textsc{JADE} \citep{naik2009effective} is object-sensitive, breaking down the problem into several context-sensitive subtasks, focusing on deadlocks between two threads in Java. Kroening et al. (\citeyear{kroening2016sound}) propose a detector for Pthread APIs that performs context-sensitive and thread-sensitive lockset analysis for lock graph construction and cycle optimization. However, exhaustive context-sensitive lockset analysis and lock graph construction are costly. \textsc{Peahen} \citep{cai2022peahen} consists of two cooperative analysis stages: context-insensitive lock graph construction and three precise yet lazy refinements, making detection more scalable and precise through context-reduction techniques.

Only a limited amount of work has focused on communication deadlocks despite the widespread use of condition variables in real-world applications. Among them, Agarwal et al. (\citeyear{agarwal2006run}) detect lost signals by checking if a wait operation always occurs before the corresponding signal event. However, they do not consider deadlocks caused by the interaction between condition variables and locks.
\textsc{FindBugs} \citep{hovemeyer2004finding} issues an alert when a thread waits on a condition variable while holding multiple locks, even though this situation does not lead to a deadlock. This oversimplified detection pattern results in a high number of false positives.
\textsc{CheckMate} \citep{joshi2010effective} records the trace of synchronization operations with manual annotations and uses a model checker to explore all possible deadlocks.  
However, the model-checking approach incurs significant runtime overhead when handling large-scale concurrent programs, and its reliance on manual annotations increases the user's burden.
\textsc{Unhang} \citep{zhou2022deadlock} is a state-of-the-art communication deadlock detection technology implemented as a dynamic library for C/C++ applications with Pthreads. It abstracts the signal for condition variables as a special resource and extends classic lock dependencies to generalized dependencies, modeling communication deadlocks as hold-and-wait cycles. This dynamic technique intercepts synchronization operations on \verb|mutex| and condition variables during execution to record generalized dependencies. Like all dynamic tools, \textsc{Unhang} cannot predict deadlocks if the relevant synchronization events do not occur at runtime. Notably, \textsc{RcChecker} and \textsc{Unhang} share similarities in the abstraction of condition variables, but \textsc{RcChecker} operates by inspecting program source code to record dependencies between locks and condition variables without relying on program execution, effectively analyzing obscure program paths. Furthermore, \textsc{RcChecker} is specifically designed for the Rust programming language, differing in several aspects. For instance, \textsc{RcChecker} leverages lifetime inference, rather than relying on explicit unlock functions, to determine the usage state of locks.

\subsection{Vulnerability Detection in Rust}
 Rust utilizes the ownership mechanism to implement automatic memory management, achieving the design goals of memory and concurrency safety, and has been applied to a range of foundational software \citep{tock,tikv,servo,loom}. However, empirical studies \citep{Boqin2020understanding, Hu2022comprehensiveness,yu2019fearless} show that Rust still has many security issues. As Rust becomes more widely used, understanding and researching its security is becoming a new focus of study. Among these, vulnerability detection has become an essential area of research in Rust language security. Based on the types of vulnerabilities, research can be divided into memory and concurrency vulnerability detection, with program analysis being the most mainstream Rust vulnerability detection technique.

\textsc{SafeDrop} \citep{cui2023safedrop} is a path-sensitive static data flow analysis method based on MIR for detecting memory release errors in Rust programs, but it has a high false positive rate when dealing with large-scale programs. And \textsc{SafeDrop} cannot collect all alias relationships, leading to false negatives. \textsc{Rupair} \citep{hua2021rupair} is a method based on MIR and AST for detecting and fixing buffer overflow vulnerabilities caused by integer operations. The aforementioned studies focus on specific types of Rust memory safety vulnerabilities. In contrast, \textsc{MirChecker} \citep{li2021mirchecker} is a vulnerability detection framework that can simultaneously detect runtime crashes and memory safety vulnerabilities caused by dangling pointers, but it does not support many advanced features of Rust, resulting in a high false positive rate. \textsc{Rudra} \citep{bae2021rudra} is a static analyzer based on MIR and HIR that uses data flow analysis algorithms and \verb|send/sync| differential check algorithms to detect memory safety vulnerabilities related to panic, high-order invariants, and \verb|send/sync| in unsafe code, but it does not support inter-procedural analysis and cannot detect memory safety vulnerabilities caused by inter-procedural interactions.

\textsc{Stuck-me-not} \citep{ning2020stuck} is a static concurrency error detector based on MIR, solely focusing on detecting double locking in blockchain software. \textsc{Lockbud} \citep{lockbud,Boqin2020understanding,qin2024understanding} is an open-source static analysis tool for detecting concurrency errors and partial memory errors. \textsc{Lockbud} utilizes type information in parameters to guide heuristic analysis in inter-procedural methods, but it suffers from numerous false positives and false negatives. In terms of deadlock detection, it mainly considers resource deadlocks. Compared to \textsc{Lockbud}, \textsc{RcChecker} supports more deadlock detection types and reduces false positives and negatives.

\section{Conclusion} \label{section7}
This paper introduces \textsc{RcChecker}, a practical static deadlock detection technique designed to achieve comprehensive detection by simultaneously identifying resource and communication deadlocks in Rust programs. Its core mechanism involves using the signal-lock graph to characterize the interactions between threads through lock variables and condition variables, effectively detecting deadlocks. \textsc{RcChecker} finds known deadlocks and reports previously undiscovered ones, demonstrating its effectiveness. We plan to employ path-sensitive techniques to improve detection precision and extend \textsc{RcChecker} to other communication synchronization mechanisms.

\backmatter

\bmhead{Acknowledgements}
This work was supported by the CCF-Huawei Populus Grove Fund (No. FM202305) and the National Natural Science Foundation of China (No. 62172299).

\section*{Declarations}

\begin{itemize}
\item Competing Interests: The authors declare no competing interests.
% \item Ethics approval and consent to participate
% \item Consent for publication
% \item Materials availability
\end{itemize}

\bibliography{sn-bibliography}% common bib file

\end{document}